\documentclass[%
 reprint,
superscriptaddress,
 amsmath,amssymb,
 aps, prx,
]{revtex4-2}

\usepackage{graphicx}
\usepackage{dcolumn}
\usepackage{bm}
\usepackage[colorlinks=true, linkcolor=red, urlcolor=blue, citecolor=red]{hyperref}
\usepackage{braket}
\usepackage{stackengine}
\usepackage{enumitem}
\usepackage{subfigure}

\begin{document}

\pagecolor{white}

\preprint{APS/123-QED}

\title{Topological Uncertainty and Higher-Order Interactions in Spatial Networks}

\author{Domenico Pomarico}
\affiliation{Dipartimento di Fisica, Università degli Studi di Bari Aldo Moro, I-70126 Bari, Italy}
\affiliation{Istituto Nazionale di Fisica Nucleare, Sezione di Bari, I-70125 Bari, Italy}

\author{Alessandro Fania}
\affiliation{Dipartimento di Fisica, Università degli Studi di Bari Aldo Moro, I-70126 Bari, Italy}
\affiliation{Istituto Nazionale di Fisica Nucleare, Sezione di Bari, I-70125 Bari, Italy}

\author{Gabriel Ramirez Sanchez}
\affiliation{Dipartimento di Fisica, Università degli Studi di Bari Aldo Moro, I-70126 Bari, Italy}

\author{Loredana Bellantuono}
\affiliation{Istituto Nazionale di Fisica Nucleare, Sezione di Bari, I-70125 Bari, Italy}
\affiliation{Dipartimento di Biomedicina Traslazionale e Neuroscienze (DiBraiN), Università degli Studi di Bari Aldo Moro, I-70124 Bari, Italy}

\author{Domenico Capolongo}
\affiliation{Dipartimento di Scienze della Terra e Geoambientali (DISTEGEO), Universit\`a degli Studi di Bari Aldo Moro, Bari, I-70125, Italy}

\author{Roberto Cilli}
\affiliation{Université de Lorraine, CNRS, GeoRessources, F-54000 Nancy, France}

\author{Alessandra Costantino}
\affiliation{Dipartimento di Fisica, Università degli Studi di Bari Aldo Moro, I-70126 Bari, Italy}

\author{Davide D'Alò}
\affiliation{Dipartimento Di Scienze Del Suolo, Della Pianta e Degli Alimenti (DISSPA), Universit\`a degli Studi di Bari Aldo Moro, Bari, I-70125, Italy}

\author{Mario Elia}
\affiliation{Dipartimento Di Scienze Del Suolo, Della Pianta e Degli Alimenti (DISSPA), Universit\`a degli Studi di Bari Aldo Moro, Bari, I-70125, Italy}

\author{Francesco Giordano}
\affiliation{Dipartimento di Fisica, Università degli Studi di Bari Aldo Moro, I-70126 Bari, Italy}
\affiliation{Istituto Nazionale di Fisica Nucleare, Sezione di Bari, I-70125 Bari, Italy}

\author{Niloofar Kheirkhahan}
\affiliation{Dipartimento di Fisica, Università degli Studi di Bari Aldo Moro, I-70126 Bari, Italy}

\author{Raffaele Lafortezza}
\affiliation{Dipartimento Di Scienze Del Suolo, Della Pianta e Degli Alimenti (DISSPA), Universit\`a degli Studi di Bari Aldo Moro, Bari, I-70125, Italy}

\author{Raffaele Nutricato}
\affiliation{GAP s.r.l., c/o Dipartimento Interuniversitario di Fisica, Universit\`a degli Studi di Bari Aldo Moro, Bari, I-70126, Italy}

\author{Ester Pantaleo}
\email{ester.pantaleo@uniba.it}
\affiliation{Dipartimento di Fisica, Università degli Studi di Bari Aldo Moro, I-70126 Bari, Italy}
\affiliation{Istituto Nazionale di Fisica Nucleare, Sezione di Bari, I-70125 Bari, Italy}

\author{Sabina Tangaro}
\affiliation{Istituto Nazionale di Fisica Nucleare, Sezione di Bari, I-70125 Bari, Italy}
\affiliation{Dipartimento Di Scienze Del Suolo, Della Pianta e Degli Alimenti (DISSPA), Universit\`a degli Studi di Bari Aldo Moro, Bari, I-70125, Italy}

\author{Roberto Bellotti}
\affiliation{Dipartimento di Fisica, Università degli Studi di Bari Aldo Moro, I-70126 Bari, Italy}
\affiliation{Istituto Nazionale di Fisica Nucleare, Sezione di Bari, I-70125 Bari, Italy}

\author{Sebastiano Stramaglia}
\affiliation{Dipartimento di Fisica, Università degli Studi di Bari Aldo Moro, I-70126 Bari, Italy}
\affiliation{Istituto Nazionale di Fisica Nucleare, Sezione di Bari, I-70125 Bari, Italy}

\author{Alfonso Monaco$^\dagger$}
\affiliation{Dipartimento di Fisica, Università degli Studi di Bari Aldo Moro, I-70126 Bari, Italy}
\affiliation{Istituto Nazionale di Fisica Nucleare, Sezione di Bari, I-70125 Bari, Italy}
\altaffiliation{these authors contributed equally to this work}

\author{Nicola Amoroso$^\dagger$}
\affiliation{Istituto Nazionale di Fisica Nucleare, Sezione di Bari, I-70125 Bari, Italy}
\affiliation{Dipartimento di Farmacia - Scienze del Farmaco, Università degli Studi di Bari Aldo Moro, I-70125 Bari, Italy}
\altaffiliation{these authors contributed equally to this work}

\date{\today}

\begin{abstract}

Environmental systems are characterized by complex spatial interactions that cannot be fully described through pairwise relationships or local uncertainty measures. We propose a unified framework combining higher-order information theory and topological data analysis to characterize the organization and uncertainty of environmental networks. Spatial entities, represented by monitoring stations or municipalities, are embedded into a Delaunay simplicial complex, and O-information is used to quantify redundancy and synergy among neighboring triplets. The resulting field of higher-order interactions is analyzed through persistent homology, enabling the identification of topological structures that remain stable across interaction scales. The methodology is applied to both an air-quality monitoring network based on weekly \(\mathrm{NO_2}\) and \(\mathrm{O_3}\) observations and a multi-hazard territorial assessment. We show that regions exhibiting strong O-information and persistent topological structures correspond to robust environmental patterns, whereas areas characterized by heterogeneous regimes and rapidly varying interactions display increased uncertainty. Building on these results, we introduce a topological uncertainty framework that integrates simplex divergence, higher-order interactions, and topological uncertainty. Our results demonstrate that uncertainty can be interpreted not only as statistical variability but also as the instability of the underlying information topology. By integrating O-information and persistent homology within a common spatial framework, the proposed approach provides a new methodology for detecting robust higher-order structures and topologically uncertain regions in environmental and multi-hazard systems.

\end{abstract}

\maketitle

\onecolumngrid



\section{Introduction}


Environmental and socio-ecological systems are increasingly characterized by complex interactions occurring across multiple spatial and temporal scales. Examples include air-quality monitoring networks (\cite{fania2024xai,fania2024airpollution}), climate-sensitive infrastructures (\cite{cilli2022, pomarico2026tensornetworkmachinelearning}), and multi-hazard landscapes (\cite{wildfire2026}) where natural and anthropogenic hazards coexist and interact. The spatial organization of such systems cannot be fully understood through conventional pairwise analyses, since important information may emerge only from the collective behavior of groups of locations rather than from individual sites or pairwise relationships. 

Recent advances in information theory have highlighted the importance of higher-order interactions in complex systems. In this context, the O-information (\cite{rosas2019oinfo, mediano2021towards, bertschinger2014quantifying, technologies13100438}) has emerged as a powerful framework for quantifying the balance between redundancy and synergy in multivariate systems. Positive O-information indicates that the variables share redundant information, whereas negative values reveal synergistic structures that only become apparent when all variables are considered jointly. Unlike pairwise mutual information, O-information provides direct access to genuinely higher-order statistical dependencies and has been successfully applied to neuroscience (\cite{Camino-Pontes2025}), genomics (\cite{Antonio_2026}), climate science, and complex networks (\cite{faes2025}). 

In parallel, topological data analysis (TDA) (\cite{lee1980twodimensional, munkres1984elements, okabe2000spatial,revTDA}) has provided a new mathematical language for studying the shape of data and it is acquiring an increasingly important role in Earth observation and geology (\cite{osti_10401888, geoTDA}). Persistent homology (\cite{edelsbrunner2002topological, zomorodian2005computing, Rote2006, carlsson2009topology, edelsbrunner2010computational, otter2017roadmap}), one of the central tools of TDA, characterizes the evolution of topological features across a filtration of simplicial complexes. By tracking the birth and death of connected components, loops, and higher-dimensional structures, persistent homology allows the identification of robust organizational patterns that persist across multiple scales while filtering out topological noise (\cite{petri2014, chintakunta2015entropy, rucco2016persistent, atienza2020stability}). 

The combination of information-theoretic and topological approaches offers an attractive framework for studying spatial systems. Information-theoretic quantities reveal the strength and nature of statistical interactions, whereas topological methods characterize the stability of the resulting spatial organization. Despite their complementary nature, these two perspectives have rarely been integrated into a unified methodology. 

We propose a topological-information framework for the analysis of environmental networks and multi-hazard systems. Spatial entities, such as monitoring stations or municipalities, are represented as vertices of a Delaunay simplicial complex constructed from their geographical coordinates. For each triangular simplex, an O-information value is computed from either pollutant time series or multi-hazard profiles, resulting in a spatial field of higher-order interactions. The resulting simplicial complex is subsequently filtered according to the magnitude of the O-information, yielding a family of nested complexes whose topological evolution can be investigated using persistent homology (\cite{simplextree}). 

Beyond the characterization of redundancy and synergy, we introduce a topological perspective on uncertainty. Traditional uncertainty measures typically quantify variability, entropy, or prediction errors. Here, uncertainty is interpreted as a lack of stable spatial organization. In this view, persistent topological structures correspond to robust and reliable patterns, whereas short-lived topological features indicate fragile configurations that are highly sensitive to perturbations. This perspective naturally connects uncertainty to the persistence of topological structures observed across multiple scales. 

To quantify such effects, we combine three complementary sources of information: (i) simplex divergence, which captures local heterogeneity of environmental regimes; (ii) O-information, which measures higher-order interactions; and (iii) persistence-based descriptors, which characterize topological robustness. The resulting framework provides both local indicators, associated with individual simplices, and global measures derived from persistence diagrams. 

The proposed methodology is applied to two representative case studies. The first concerns a national air-quality monitoring network based on weekly observations, where Delaunay simplices connect neighboring monitoring stations. Air pollution represents a particularly relevant application of this framework, as pollutant concentrations result from the combined influence of emission sources, atmospheric transport, meteorological conditions, and chemical transformations. Among the major pollutants of public-health concern, nitrogen dioxide (\(\mathrm{NO_2})\) is mainly associated with combustion processes, whereas ground-level ozone (\(\mathrm{O_3})\) is a secondary pollutant formed through photochemical reactions (\cite{who2021airquality}). Previous studies integrated monitoring-station observations with satellite, meteorological, and geographical data to reconstruct spatially continuous pollutant distributions and investigate their relevance within a One Health perspective (\cite{fania2024xai,fania2024airpollution}). The second case study involves a spatial multi-hazard assessment (\cite{wildfire2026}) in which municipalities are represented as nodes and characterized by multiple susceptibility indicators. Higher-order analyses are essential in multi-hazard assessment because environmental hazards rarely act independently, and their combined effects often generate compound or cascading processes that cannot be captured by pairwise relationships alone. In this perspective, the integration of higher-order information measures with topological descriptors provides a multiscale framework for investigating how complex hazard interactions organize and persist across space. 

The main contribution of this work is therefore the integration of O-information and persistent homology into a common spatial framework, providing a novel interpretation of uncertainty as a property of the topology of information interactions. This approach enables the identification of stable environmental structures, synergistic regions, and areas characterized by fragmented or poorly organized spatial behavior, thereby offering a new perspective for the analysis of complex environmental systems.

\section{Materials}


Two spatial datasets were considered in this study to investigate higher-order interactions and topological uncertainty in environmental systems. The first dataset consists of an air-quality monitoring network based on weekly observations of nitrogen dioxide (\(\mathrm{NO_2}\)) and ozone (\(\mathrm{O_3}\)) collected at monitoring stations distributed across Italy. The second dataset covers the Province of Bari and the Gargano area in Apulia (Southern Italy) and consists of a multi-hazard spatial framework in which municipalities are characterized by multiple environmental susceptibility indicators (\cite{wildfire2026}), i.e. wildfires, floods, land displacements and sediment connectivity, represented as raster layers. 

\subsection{Air-Quality Monitoring Network} 

The air-quality dataset consists of weekly observations of nitrogen dioxide (\(\mathrm{NO_2}\)) and ozone (\(\mathrm{O_3}\)) collected from a network of monitoring stations distributed across Italy. 

Hourly observations were retrieved from the European Environment Agency Air Quality Download Service for the period from January 2025 to December 2025 (\cite{eeaairquality}). All concentrations were converted to (\( \mu\mathrm{g}/\mathrm{m}^{3}\)) and aggregated into Monday-to-Sunday weekly means. Before aggregation, missing or non-finite observations, negative values, including negative sentinel codes, and measurements associated with invalid EEA quality flags were excluded. Duplicate records were removed, and when multiple sampling-point codes were available for the same station and pollutant, the sampling point with the largest overall hourly coverage was selected. A weekly concentration was retained only when at least 75\% of the 168 expected hourly observations were available. Partial weeks at the beginning and end of the study period were excluded, and each station--week record was included in the final dataset only when valid weekly concentrations of both \(\mathrm{NO_2}\) and \(\mathrm{O_3}\) were simultaneously available.

Each station is characterized by its geographical coordinates and by a multivariate time series describing the temporal evolution of pollutant concentrations. Let \( S_i \) denote the \(i\)-th monitoring station and let \( \mathcal{T}_i = \{t_1,t_2,\ldots,t_{N_i}\} \) represent the set of observation weeks available at that station. For every observation time \(t\in\mathcal{T}_i\), a two-dimensional pollutant vector is defined as 
\begin{equation}
\mathbf{x}_i(t) = \Big( \mathrm{NO}_2(t), \mathrm{O}_3(t) \Big). 
\end{equation}
Consequently, each monitoring station is represented by the collection of all observed pollutant vectors, \( X_i = \left\{ \mathbf{x}_i(t) : t\in\mathcal{T}_i \right\}, \) which can be expressed in matrix form as \( X_i \in \mathbb{R}^{N_i \times 2}, \) where \(N_i\) denotes the number of weekly observations available for station \(S_i\). The rows correspond to observation weeks and the two columns correspond to \(\mathrm{NO_2}\) and \(\mathrm{O_3}\), respectively. Rather than characterizing a station through aggregated statistics such as mean concentrations, the proposed framework exploits the complete multivariate pollutant trajectory observed at each location. This representation preserves temporal variability, seasonal fluctuations, and the joint behavior of the two pollutants, allowing higher-order information-theoretic quantities to be estimated directly from the observed distributions. To compare neighboring stations, weekly observations were temporally aligned. Prior to the information-theoretic analysis, each station-specific matrix was standardized to zero mean and unit variance in order to remove scale effects associated with different concentration ranges and to focus on the intrinsic dependency structure of the pollutant dynamics. The geographical coordinates of all monitoring stations were used.

\begin{figure}
    \centering
    \subfigure[]{\includegraphics[width=0.9\linewidth]{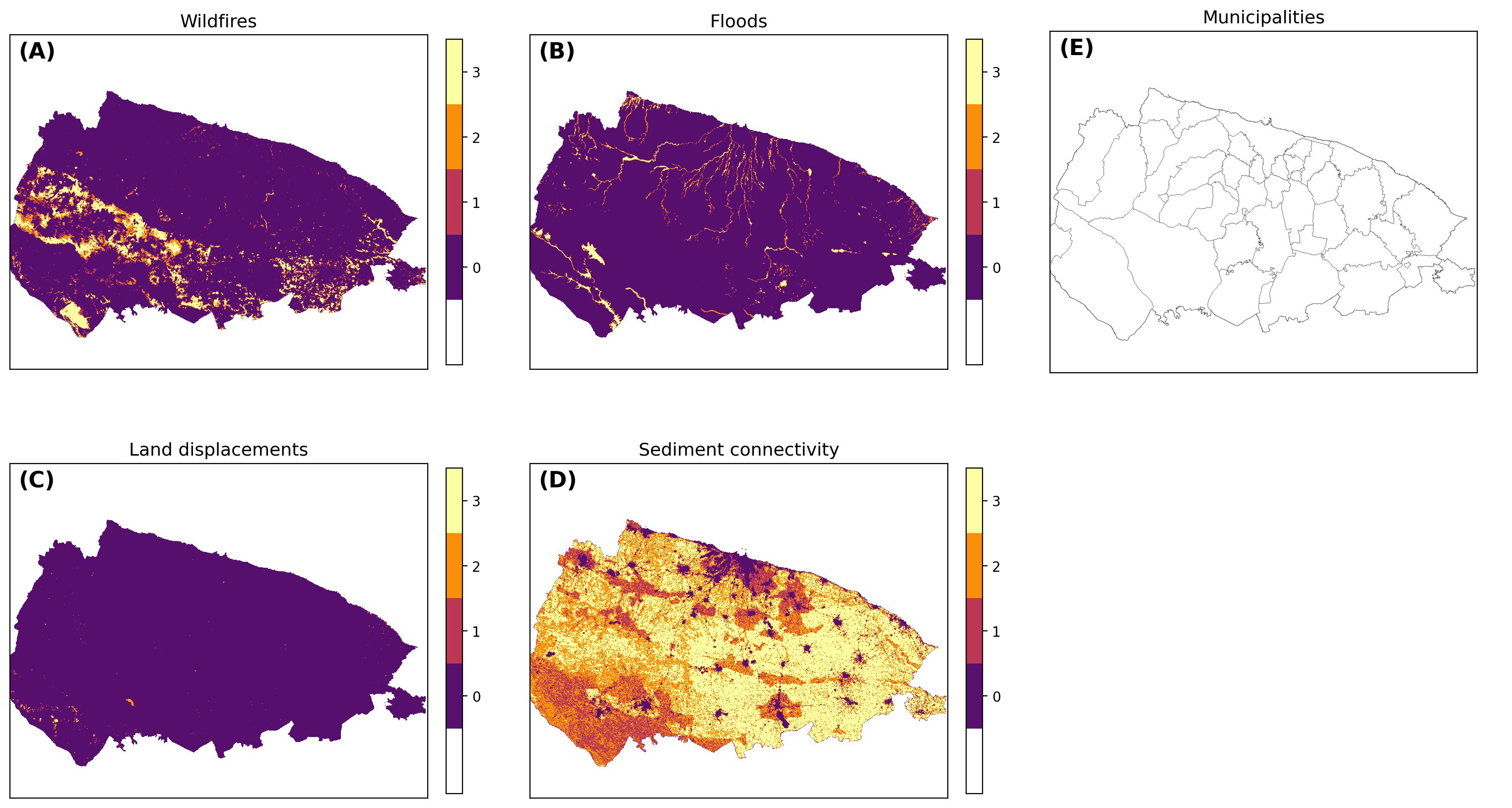}}
        \caption{Province of Bari.}
    \subfigure[]{\includegraphics[width=0.9\linewidth]{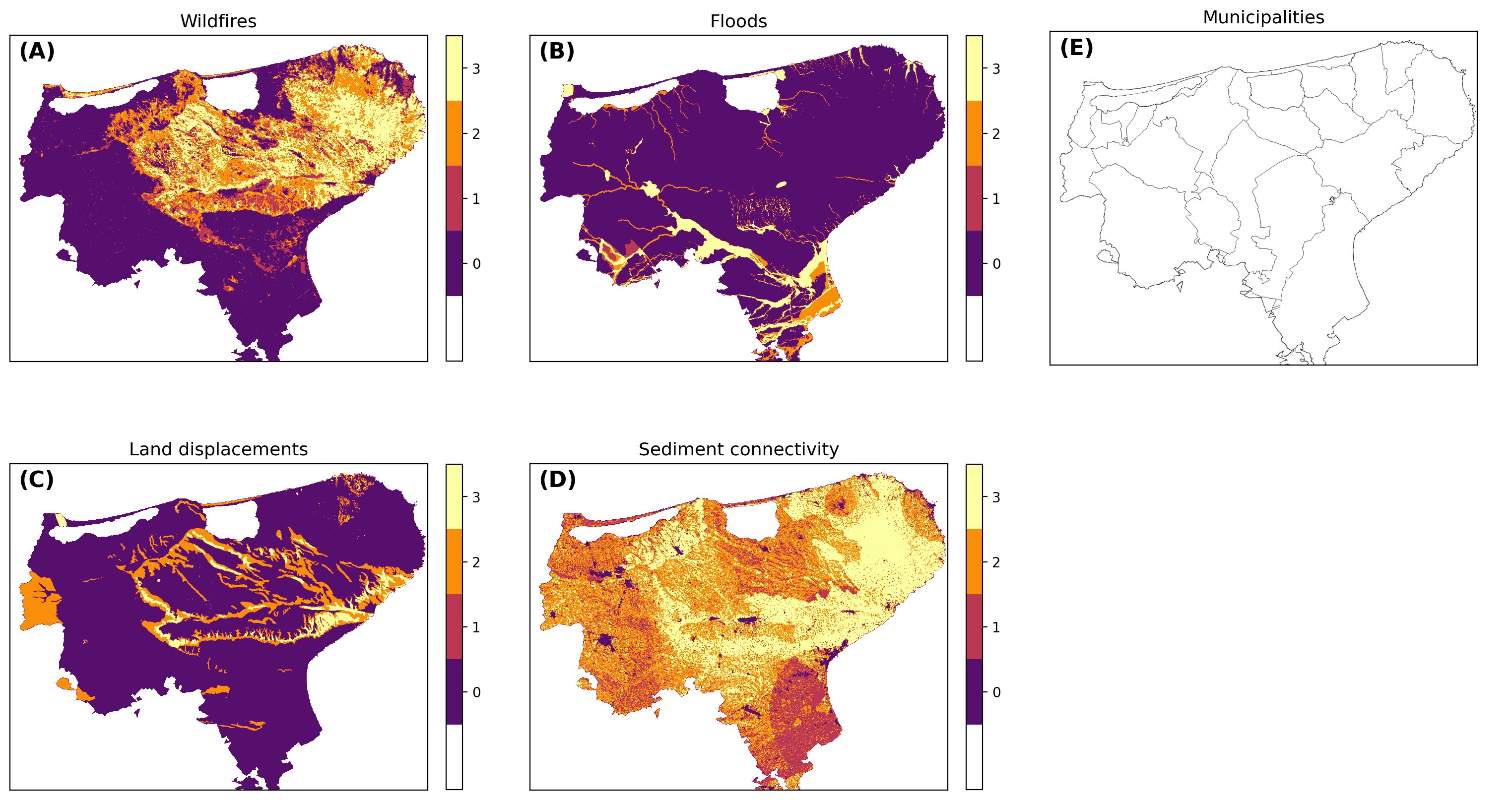}}
        \caption{Gargano area.}
    \caption{Multi-hazard layers and municipality partitioning for the Province of Bari \textbf{(a)} and Gargano area \textbf{(b)}. Panels (A)-(D) depict wildfire, flood, land-displacement, and sediment-connectivity susceptibility layers, respectively. Panels (E) show the municipality labeling map used to aggregate pixel-level susceptibility information.}
    \label{fig:multirisk}
\end{figure}

\subsection{Multi-hazard Spatial Dataset} 

The multi-hazard dataset (\cite{wildfire2026}) consists of four raster layers, depicted in Figure \ref{fig:multirisk}, representing distinct environmental susceptibility components together with an administrative raster identifying municipality boundaries, whose pixels correspond to an area $10 \ m \times 10 \ m$.
Table \ref{tab:environmental_datasets}, summarizes the datasets used to produce the four susceptibility maps, together with their corresponding reference years. The flood-hazard map was originally classified into four hazard levels. To ensure consistency across indicators, the wildfire, sediment, and land-displacement maps were likewise discretized into four ordinal classes. The complete data-processing workflow used to derive the final susceptibility maps is described in detail by (\citet{wildfire2026}).

\begin{table}[!t]
    \centering
    \caption{Summary of environmental hazard and connectivity datasets.}
    \label{tab:environmental_datasets}
    \renewcommand{\arraystretch}{1.4}
    \begin{tabular}{p{2.8cm}|p{11cm}}
        \hline
        \textbf{Dataset} &
        \textbf{Description and data source} \\
        \hline
        Floods &
        Data derived from the Flood Risk Management Plan of the River
        Basin Authority-Southern Apennines District, corresponding
        to the 2016-2021 planning cycle (\cite{adbdam_ii_ciclo}). \\
        Land displacements &
        Combination of the Italian geomorphological hazard map and spatial
        anomalies identified from Persistent and Distributed Scatterers using
        velocity thresholds and spatial clustering criteria. Data were derived
        from the Italian geomorphological hazard map produced by the Italian
        Institute for Environmental Protection and Research and local
        Hydrogeological Management Plans. Data updated to 2024 (\cite{trigila2025dissesto}).\\
        Sediment connectivity &
        Map of sediment connectivity derived from post-processing the Sediment
        Flow Connectivity Index (SFCI), a geomorphological indicator of surface
        sediment mobility and flow within a watershed.\\
        Wildfires &
        Wildfire hazard obtained by combining the hazard component, representing
        the potential presence of forest-fuel elements that predispose ignition,
        and the danger component, representing the probability of ignition under
        predisposing conditions. \\
        \hline
    \end{tabular}
\end{table}
Let \( \mathcal{S}_s(x,y), s=1,\ldots,4, \) denote the four susceptibility layers defined on a regular spatial grid, and let \( M(x,y) \) be the municipality label raster, where each pixel is assigned an integer identifier corresponding to the municipality to which it belongs. Pixels with non-administrative labels (e.g., background or external areas) were excluded from the analysis. For a municipality \(M_i\), all pixels satisfying \( M(x,y)=i \) were extracted, yielding the set 
\begin{equation}
    \mathcal{P}_i = \left\{ (x_p,y_p): M(x_p,y_p)=i \right\}. 
\end{equation}
Each pixel \(p \in \mathcal{P}_i\) is characterized by a four-dimensional multi-hazard vector 
\begin{equation}
\mathbf{x}_p = \Big( \mathcal{S}_1(p), \mathcal{S}_2(p), \mathcal{S}_3(p), \mathcal{S}_4(p) \Big), 
\end{equation}
which encodes the local intensity of the four susceptibility components at that spatial location. Consequently, municipality \(M_i\) is represented not by a single summary statistic but by the collection of all susceptibility vectors associated with its pixels: \( X_i = \left\{ \mathbf{x}_p : p \in \mathcal{P}_i \right\}. \) In matrix form, \( X_i \in \mathbb{R}^{N_i \times 4}, \) where \(N_i\) denotes the number of pixels belonging to municipality \(M_i\), where each row corresponds to a pixel and each column corresponds to one of the four susceptibility variables. This representation preserves both the spatial variability and the multivariate structure of the susceptibilities within each municipality. Rather than comparing municipalities through aggregated averages alone, the proposed framework exploits the full distribution of multi-hazard pixel values contained within each administrative unit. As a result, the subsequent information-theoretic analysis captures higher-order relationships among the complete susceptibility distributions of neighboring municipalities. Municipality centroids \( \mathbf{c}_i=(x_i,y_i) \) were computed from the corresponding administrative masks.

\section{Methods}

Environmental and multi-hazard systems are characterized by multiple interacting processes operating across different spatial scales. To investigate this complexity, we combine information-theoretic and topological approaches within a common simplicial framework. Local heterogeneity is quantified through a simplex divergence, which extends the pairwise comparison of probability distributions to multiple vertices of a simplex, i.e. a quantity measuring differences in environmental and multi-hazard compositions. Higher-order interactions are characterized through the O-information, a multivariate measure that quantifies whether neighboring environmental variables are organized through redundancy or synergy. Positive values characterize redundancy-dominated information sharing, while negative values characterize synergy-dominated information integration that emerge only from the joint contribution of all variables. Persistent homology is employed to investigate the stability of the resulting interaction patterns. By progressively thresholding weighted graphs and simplicial complexes derived from the O-information field, a sequence of binarized topological structures is generated. Persistent homology tracks the appearance and disappearance of connected components, loops, and higher-order features across the threshold range, allowing robust environmental structures to be distinguished from short-lived fluctuations. In this way, topological persistence provides a multiscale description of the resilience of environmental and multi-hazard interaction patterns under varying levels of interaction strength.

\subsection{Simplex Divergence}

The heterogeneity of the environmental regimes represented within a simplex is captured by the simplex divergence measure that quantifies the diversity of the categorical labels associated with the vertices of each triangular simplex. Let \( \tau=(v_i,v_j,v_k) \) be a simplex of the Delaunay simplicial complex, yielding vertices, edges and triangular simplices connecting neighboring locations while preserving the local geometry of the spatial network. Each vertex is associated with an empirical probability distribution describing the local environmental composition or atmospheric regime observed at the corresponding spatial unit.

In the multi-hazard framework, the four susceptibility layers consist of categorical susceptibility labels defined at pixel level. For each municipality \(M_i\), all pixels belonging to the municipality are extracted and the empirical distribution of susceptibility labels is computed separately for every susceptibility category. Therefore, municipality \(M_i\) is represented by a collection of probability vectors 
\begin{equation} \label{eq:class_prob}
    P_i^{(s)} = \left( p_{i1}^{(s)}, p_{i2}^{(s)}, p_{i3}^{(s)}, p_{i4}^{(s)} \right), 
\end{equation}
where \(p_{ic}^{(s)}\) denotes the fraction of municipality pixels assigned to class \(c\) for susceptibility layer \(s\). This representation preserves the full internal composition of the municipality rather than reducing it to a single dominant susceptibility label. In this way the probability distributions employed in the Jensen--Shannon divergence analysis are not obtained through histogram binning, kernel density estimation, or other density reconstruction procedures. For a given municipality, the relative frequencies of the risk classes are computed separately for each hazard, yielding an empirical probability distribution without the introduction of additional smoothing or discretization parameters. This choice can be interpreted as a coarse-graining strategy that shifts the analysis from the pixel scale to the municipality scale, thereby preserving the dominant structure of the risk landscape while reducing the influence of local fluctuations and isolated pixel-level anomalies. In contrast to histogram-based or kernel-based approaches, which require the specification of bin widths, bandwidths, or normalization schemes and may therefore introduce methodological arbitrariness, the adopted formulation produces probability distributions directly determined by the observed risk-class frequencies. Consequently, the resulting Jensen--Shannon divergence quantifies differences in municipality-level risk compositions through a parameter-free and physically interpretable representation, allowing meaningful comparisons among heterogeneous environmental risk indicators.

For the air-quality monitoring network, each station is characterized by the temporal evolution of the ratio \( R_i(t) = \frac{\mathrm{NO}_2(t)} {\mathrm{O}_3(t)}, \) computed for every observation week. The empirical distribution of ratio values across all stations and weeks is partitioned into four quartile-based classes, ranging from strongly ozone-dominated to nitrogen-dioxide-dominated conditions. Each weekly observation is therefore assigned to one of four atmospheric regimes. For station \(i\), the relative frequency of occurrence of the four regimes is computed over the entire observation period, yielding the probability vector \( P_i = (p_{i1},p_{i2},p_{i3},p_{i4}), \) where \(p_{ic}\) denotes the fraction of observations of station \(i\) belonging to regime \(c\), according to an equivalent formulation with Equation (\ref{eq:class_prob}). Consequently, each station is represented by a probability distribution of pollutant states rather than by a single dominant atmospheric class.

To quantify the heterogeneity of a simplex, the distributions associated with its three vertices are compared through a Jensen-Shannon formulation (\cite{nielsen2019}). For a given susceptibility category \(s\), let \( P_i^{(s)}, P_j^{(s)}, P_k^{(s)}\) be the distributions associated with the municipalities composing simplex \(\tau\). Their barycentric mixture is 
\begin{equation}
    M^{(s)} = \frac{ P_i^{(s)} + P_j^{(s)} + P_k^{(s)} }{3}. 
\end{equation} 
The Jensen-Shannon simplex divergence contribution of susceptibility \(s\) is then defined as 
\begin{equation}
\mathrm{JS}_s(\tau) = H\!\left(M^{(s)}\right) - \frac{1}{3} \Big[ H\!\left(P_i^{(s)}\right) + H\!\left(P_j^{(s)}\right) + H\!\left(P_k^{(s)}\right) \Big], 
\end{equation}  
where \( H(P) = -\sum_c p_c \log p_c \) denotes the Shannon entropy \cite{shannon1948mathematical}. The final simplex divergence is obtained by averaging the contributions of all susceptibility categories and normalizing the result to the interval \([0,1]\). Low simplex divergence indicates that neighboring municipalities or stations share similar environmental compositions and therefore belong to a coherent local regime. Conversely, high simplex divergence identifies transition regions where different environmental states coexist within the same neighborhood. In this sense, simplex divergence acts as a local configurational descriptor of environmental heterogeneity, or equivalently simplex divergence quantifies the diversity of the environmental regimes represented within a simplex. Consequently, a simplex may exhibit strong higher-order interactions while remaining compositionally homogeneous, or vice versa. Within the proposed framework, simplex divergence therefore provides complementary information that links local environmental diversity to higher-order spatial organization.

In the air-quality application, each station is represented by a single probability distribution \( P_i, \) describing the relative frequencies of the four NO$_2$/O$_3$ atmospheric regimes observed throughout the monitoring period. Consequently, only one distribution is associated with each vertex and no averaging across susceptibility categories is required. For a simplex \( \tau=(i,j,k), \) the simplex divergence is therefore directly computed through the Jensen--Shannon divergence 
\begin{equation}
    \mathrm{JS}(\tau) = H(M) - \frac{1}{3} \Big[ H(P_i) + H(P_j) + H(P_k) \Big], 
\end{equation}
where \( M=\frac{P_i+P_j+P_k}{3}\). The resulting quantity measures the dissimilarity among the temporal pollutant-regime distributions of the three stations composing the simplex. Values close to zero indicate stations exhibiting similar atmospheric behavior over time, whereas large values identify neighboring stations characterized by markedly different pollutant-regime compositions.

\subsection{O-Information Evaluation} 

Higher-order interactions among neighboring spatial entities are quantified through the O-information, a multivariate information-theoretic measure capable of distinguishing redundant and synergistic dependencies (\cite{rosas2019oinfo,Camino-Pontes2025,technologies13100438}). For three variables, the O-information coincides with the interaction information and can be expressed as 
\begin{equation} \Omega(X,Y,Z) = I(X;Y) - I(X;Y|Z), \label{eq:o_information} 
\end{equation} 
where \(I(X;Y)\) denotes the mutual information between \(X\) and \(Y\), and \(I(X;Y|Z)\) is the corresponding conditional mutual information. Positive values of \(\Omega\) indicate redundancy-dominated interactions, whereas negative values reveal synergistic dependencies that can only be explained by considering the three variables jointly. For the air-quality monitoring network, each Delaunay simplex \( \tau=(S_i,S_j,S_k) \) is associated with three multivariate pollutant processes \(X_i\), constructed from the weekly observations available at station \(S_i\). The common observation period \( \mathcal{T}_{ijk} = \mathcal{T}_i \cap \mathcal{T}_j \cap \mathcal{T}_k \) is first determined to guarantee temporal alignment among the three stations. The corresponding matrices \( X_i^{(ijk)}, X_j^{(ijk)}, X_k^{(ijk)} \) are subsequently standardized to zero mean and unit variance before the information-theoretic analysis. 

For the multi-hazard framework, each municipality \(M_i\) is represented by the ensemble of pixel-level multivariate susceptibility vectors \(X_i\) belonging to the corresponding administrative unit. For a simplex \( \tau=(M_i,M_j,M_k), \) the variables entering Equation~(\ref{eq:o_information}) therefore correspond to empirical multivariate distributions rather than aggregated municipality-level statistics. For numerical consistency, the three municipality datasets are restricted to the same number of samples, corresponding to the minimum number of available pixels among the municipalities composing the simplex.

Mutual information and conditional mutual information are estimated by means of the Kraskov-Stögbauer-Grassberger (KSG) nearest-neighbor approach (\cite{kraskov2004estimating,frenzel2007partial}). Given two multivariate variables \(X\) and \(Y\), the mutual information is evaluated as
\begin{equation} 
I(X;Y) = \psi(k) + \psi(N) - \Big\langle \psi(N_X+1) + \psi(N_Y+1) \Big\rangle, \label{eq:ksg_mi} 
\end{equation} 
where \(N\) is the number of observations, \(k\) is the selected nearest-neighbor order, \(\psi(\cdot)\) denotes the digamma function, and \(N_X\) and \(N_Y\) are the numbers of neighboring points contained within the joint-space radius determined by the \(k\)-th nearest neighbor. Similarly, the conditional mutual information is estimated according to 
\begin{equation} 
I(X;Y|Z) = \psi(k) + \Big\langle \psi(N_Z+1) - \psi(N_{XZ}+1) - \psi(N_{YZ}+1) \Big\rangle, \label{eq:ksg_cmi} 
\end{equation} 
where \(N_Z\), \(N_{XZ}\), and \(N_{YZ}\) denote the corresponding neighbor counts in the marginal spaces. Distances are evaluated using the Chebyshev metric, consistently with the original KSG formulation. For every simplex of the Delaunay complex, Equations~(\ref{eq:ksg_mi}) and (\ref{eq:ksg_cmi}) are combined through Equation~(\ref{eq:o_information}) to obtain a simplex-specific O-information value \( \Omega(\tau). \) The collection of O-information values defines a scalar field over the simplicial complex, providing a spatial characterization of higher-order interactions. Redundant regions are associated with positive O-information and indicate spatial neighborhoods sharing similar information content, whereas negative values reveal synergistic configurations in which the joint behavior of the three vertices contains information that cannot be inferred from pairwise relationships alone.

\subsection{Persistent Homology, $\Omega$-Weighted Persistence and Local Robustness} 

To characterize the spatial organization of higher-order interactions, we associate each Delaunay simplicial complex with a filtration induced by the magnitude of the O-information. Let \( K=\mathrm{Del}(P) \) denote the simplicial complex generated from the geographical coordinates of monitoring stations or municipality centroids. For every triangular simplex \( \tau=(i,j,k), \) an O-information value \( \Omega(\tau) \) is computed according to Equation~(\ref{eq:o_information}). The filtration function is then defined as \( f(\tau) = |\Omega(\tau)|. \) Vertices are assigned filtration value zero, whereas each edge and triangular simplex inherit the filtration value of the corresponding O-information. The resulting filtration is given by 
\begin{equation}
    K_{\alpha} = \{ \sigma \in K : f(\sigma)\le \alpha \}, 
\end{equation} 
which defines a nested sequence of simplicial complexes \( K_{\alpha_1} \subseteq K_{\alpha_2} \) for \( \alpha_1<\alpha_2 \) (\cite{simplextree}). Persistent homology is subsequently computed on the filtered complex. For every topological feature, the persistence algorithm returns a birth–death pair \( (b_i,d_i), \) where \(b_i\) and \(d_i\) denote the filtration values at which the feature appears and disappears, respectively. The corresponding persistence length is \( \ell_i = d_i-b_i. \) Large persistence values identify topological structures that remain stable across a broad range of O-information scales, whereas short intervals are typically associated with weakly supported or highly localized structures. 

Using a fixed Delaunay complex separates spatial geometry from the O-information filtration, improving interpretability and preserving standard stability properties with respect to perturbations of the filtration function. It also yields a sparse, geographically meaningful complex, avoiding the combinatorial growth of more general constructions such as Vietoris–Rips complexes. This property improves both computational feasibility and interpretability, since the resulting simplices correspond to geographically neighboring entities with direct environmental meaning. From a theoretical perspective, the use of a fixed simplicial complex allows the persistence diagrams to inherit the standard stability properties of persistent homology with respect to perturbations of the filtration function. In particular, small variations of the O-information field produce bounded changes in the associated persistence diagrams, ensuring that the identified topological structures are not generated solely by minor fluctuations of the underlying environmental variables. The persistence analysis should therefore be interpreted as a multiscale characterization of the stability of higher-order spatial interactions on a fixed geographic network rather than as a topology induced by continuously varying spatial neighborhoods.


\subsubsection{Local persistence assignment.}

Persistence intervals are intrinsically global descriptors of the filtered simplicial complex. In order to obtain simplex-wise quantities suitable for spatial visualization, each simplex is associated with the persistence interval whose birth scale is closest to its filtration value. Specifically, for a simplex \(\tau\) with filtration value \( |\Omega(\tau)|, \) the corresponding persistence score is defined as 
\begin{equation}
    \ell(\tau) = \ell_i, \qquad i = \arg\min_j \left| b_j - |\Omega(\tau)| \right|. 
\end{equation} 
This procedure is heuristic, and it provides a local proxy of topological persistence that can be directly mapped onto the simplicial complex. 

\subsubsection{$\Omega$-weighted persistence.} 

To jointly account for the strength of higher-order interactions and the persistence of the corresponding topological structures, an \(\Omega\)-weighted persistence measure is introduced: 
\begin{equation}
    \mathrm{OWP}(\tau) = |\Omega(\tau)| \,\ell(\tau). 
\end{equation} 
The persistence length \(\ell(\tau)\) is obtained by associating simplex \(\tau\) with the persistence interval whose birth value is closest to the filtration value \(|\Omega(\tau)|\). The persistence diagram itself is generated from the simplicial filtration induced by the O-information field, where shared lower-dimensional simplices inherit the earliest admissible filtration value among the incident triangles. Consequently, \(\mathrm{OWP}\) combines the local strength of higher-order interactions with the robustness of the corresponding topological structures across the filtration. Large values of \(\mathrm{OWP}\) identify simplices characterized simultaneously by strong higher-order interactions and long-lived topological features, whereas small values indicate either weak interactions or topologically uncertain structures.

The persistence intervals are characterized by their persistence lengths \( \ell_i \). Following the persistence entropy formalism (\cite{chintakunta2015entropy, rucco2016persistent}), a probability distribution is defined as \( q_i = \frac{\ell_i} {\sum_j \ell_j}. \) The corresponding persistence entropy is 
\begin{equation}
    S_{\mathrm{pers}} = -\sum_i q_i\log q_i. 
\end{equation} 
A normalized Topological Uncertainty Index (TUI) is then obtained as 
\begin{equation}
    TUI = \frac{ S_{\mathrm{pers}} }{ \log m }, 
\end{equation} 
where \(m\) denotes the number of finite persistence intervals, and it satisfies \( 0 \le TUI \le 1 \). Large values of \(TUI\) indicate the presence of multiple competing topological scales, whereas small values reveal a topology dominated by a few persistent structures. Consequently, \(TUI\) should be interpreted as a measure of the diversity or dispersion of persistence scales rather than as a direct estimate of statistical uncertainty. In environmental applications, large \(TUI\) values correspond to persistence spectra distributed across many features with comparable lifetimes, while smaller values indicate that the persistence diagram is dominated by a small number of long-lived topological structures. It is important to note that \(TUI\) depends exclusively on the persistence lengths and therefore characterizes the topology of the filtration independently of the magnitude of the underlying O-information field. For this reason, \(TUI\) is complemented by the \(\Omega\)-weighted persistence measure, which combines topological stability and higher-order interaction strength within a single descriptor.

\subsubsection{Local robustness.} 

While persistence entropy provides a global characterization of topological organization, spatially localized robustness is evaluated directly on the O-information field. Let \( \mathcal{N}(\tau) \) denote the set of triangles sharing at least one edge with simplex \(\tau\). The local robustness score is defined as 
\begin{equation}
    R(\tau) = \frac{ |\Omega(\tau)| }{ \operatorname{Var} \!\left( \Omega(\mathcal{N}(\tau)) \right)}. 
\end{equation} 
This quantity increases when a simplex exhibits strong higher-order interactions embedded within a spatially coherent neighborhood and decreases when large fluctuations of O-information occur among adjacent simplices. The robustness scores are finally normalized to the interval \([0,1]\) through \( \widetilde{R}(\tau) = \frac{ R(\tau) }{ \max_{\tau} R(\tau) }. \) Consequently, large values of \(\widetilde{R}\) identify spatially coherent and structurally stable regions, whereas low values indicate locally heterogeneous areas characterized by weak or rapidly varying higher-order interactions.

\subsection{Spatial Autocorrelation Analysis: Local Moran, LISA Clusters and Getis-Ord Statistics}

To complement the information-theoretic and topological analyses, we evaluated the spatial organization of municipality-level multirisk compositions through local indicators of spatial association (LISA) (\cite{lisa}). Unlike O-information, which quantifies higher-order dependencies among neighboring municipalities, Moran's \(I\) (\cite{moran}) and Getis-Ord (\cite{getis-ord}) statistics explicitly assess whether municipalities characterized by pairwise similar multi-hazard compositions tend to cluster in space. Each municipality \(M_i\) is represented by a probability tensor presented in Equation \eqref{eq:class_prob}. The four probability vectors are subsequently flattened into a single municipality descriptor  \( X_i \in \mathbb{R}^{16}, \) providing a multivariate representation of the complete multi-hazard composition of municipality \(i\).  For the air-quality monitoring network, spatial autocorrelation is evaluated on the scalar station-level ratio \( R_i = \frac{\overline{\mathrm{NO}_2}} {\overline{\mathrm{O}_3}}, \) leading to the classical scalar definitions of Moran's \(I\), Local Moran's \(I\), and Getis--Ord statistics.


Spatial relationships among municipalities are defined through the Delaunay triangulation constructed from municipality centroids. In the graph induced by the triangulation, two municipalities \(i\) and \(j\) are considered neighbors if they share an edge of the Delaunay graph. The resulting adjacency matrix is 
\[ w_{ij} = \begin{cases} 1, & \text{if } i \text{ and } j \text{ are Delaunay neighbors},\\ 0, & \text{otherwise}. \end{cases} \] 
The adjacency matrix is subsequently row-normalized, \( \widetilde w_{ij} = \frac{w_{ij}} {\sum_j w_{ij}}, \) yielding a spatial weights matrix whose rows sum to unity. This normalization ensures that each municipality contributes equally to the local spatial statistics independently of the number of neighbors generated by the triangulation.


Because each municipality is described by a multivariate vector rather than a scalar quantity, we adopt a multivariate generalization of Moran's coefficient. Let \( \overline X = \frac{1}{N} \sum_{i=1}^{N} X_i \) denote the mean multirisk composition of the study area and define the centered vectors \( X_i^\prime = X_i-\overline X. \) The global multivariate Moran index is computed as
\[ I_M = \frac{N}{S_0} \frac{ \sum_{i=1}^{N} \sum_{j=1}^{N} \widetilde w_{ij} \, \left( X_i^\prime, X_j^\prime \right) } { \sum_{i=1}^{N} \|X_i^\prime\|^2 }, \]
where \( S_0 = \sum_{i,j} \widetilde w_{ij}, \) \(\left(\cdot,\cdot\right)\) denotes the Euclidean inner product, and \(\|\cdot\|\) is the Euclidean norm. Positive values indicate that municipalities characterized by similar multirisk compositions are spatially clustered, whereas values near zero are associated with weak global spatial autocorrelation.


To identify local spatial structures, a municipality-specific multivariate Moran score is computed as 
\[ I_i = \left( X_i^\prime, \sum_j \widetilde w_{ij} X_j^\prime \right). \]
Large positive values indicate municipalities embedded within neighborhoods characterized by similar multirisk compositions, while negative values identify local anomalies. Following the LISA framework, municipalities are classified into four categories. Let \( L_i = \|X_i^\prime\| \) be the magnitude of the centered multirisk vector and \( S_i = \left\| \sum_j \widetilde w_{ij} X_j^\prime \right\| \) the corresponding neighborhood magnitude. Municipalities are partitioned according to whether \(L_i\) and \(S_i\) are above or below their respective median values, yielding: 
\begin{itemize} 
\item High--High (HH): municipalities with large multirisk deviations surrounded by municipalities exhibiting similarly large deviations; 
\item Low--Low (LL): municipalities with weak multirisk deviations embedded within similarly weak neighborhoods; 
\item High--Low (HL): local multirisk outliers characterized by stronger deviations than their surroundings; 
\item Low--High (LH): municipalities with weaker deviations located within highly structured neighborhoods. 
\end{itemize} 
These classes provide a multivariate extension of the traditional LISA cluster interpretation.


To identify local concentrations of multirisk compositions, we introduce a multivariate extension of the Getis--Ord statistic. For municipality \(i\), the neighborhood composition vector is
\[ G_i = \sum_j \widetilde w_{ij} X_j. \] 
Its magnitude \( g_i = \|G_i\| \) quantifies the intensity of multirisk organization within the local neighborhood. To facilitate interpretation, the magnitudes are standardized according to \( Z_i = \frac{ g_i-\overline g } { \sigma_g }, \) where \(\overline g\) and \(\sigma_g\) denote the mean and standard deviation of the neighborhood magnitudes over all municipalities. Large positive values of \(Z_i\) identify multirisk hot-spots characterized by strong local organization, whereas negative values denote cold-spots associated with weakly structured multirisk environments. Following the conventional interpretation of standardized scores, municipalities satisfying \( Z_i > 1.96 \) are considered statistically significant hot-spots at the \(95\%\) confidence level, while \( Z_i < -1.96 \) identify statistically significant cold-spots.

\subsection{Software Environment} 

All analyses were implemented in Python. Numerical computations were performed using \texttt{NumPy}, \texttt{SciPy} and \texttt{Pandas}. Delaunay triangulations were generated using the \texttt{scipy.spatial} module. Entropic quantities were estimated through \(k\)-nearest-neighbor methods implemented using \texttt{scikit-learn}. Persistent homology computations were performed using the \texttt{Gudhi} library, while spatial visualizations were generated using \texttt{Matplotlib} and \texttt{GeoPandas}. The complete workflow includes the construction of spatial simplicial complexes, estimation of O-information on triangular simplices, persistent homology analysis of O-information filtrations, and the derivation of local and global topological uncertainty indicators.

For the air-quality monitoring network, O-information was estimated from temporally aligned \(\mathrm{NO_2}\)-\(\mathrm{O_3}\) time series. Only simplices sharing at least five common observation weeks among the three monitoring stations were retained for the analysis, and the KSG estimator was evaluated using \(k=4\) nearest neighbors. For the multi-hazard applications, O-information was computed from municipality-level datasets composed of pixelwise susceptibility vectors sampled from the four susceptibility layers. The nearest-neighbor parameter was set to \(k=400\) for the Province of Bari and \(k=200\) for the Gargano area, reflecting the different sizes and statistical sampling densities of the corresponding datasets.

\section{Results}

\subsection{Air-Quality Monitoring Network}

\begin{figure}[!t]
\begin{center}
\includegraphics[width=0.9\linewidth]{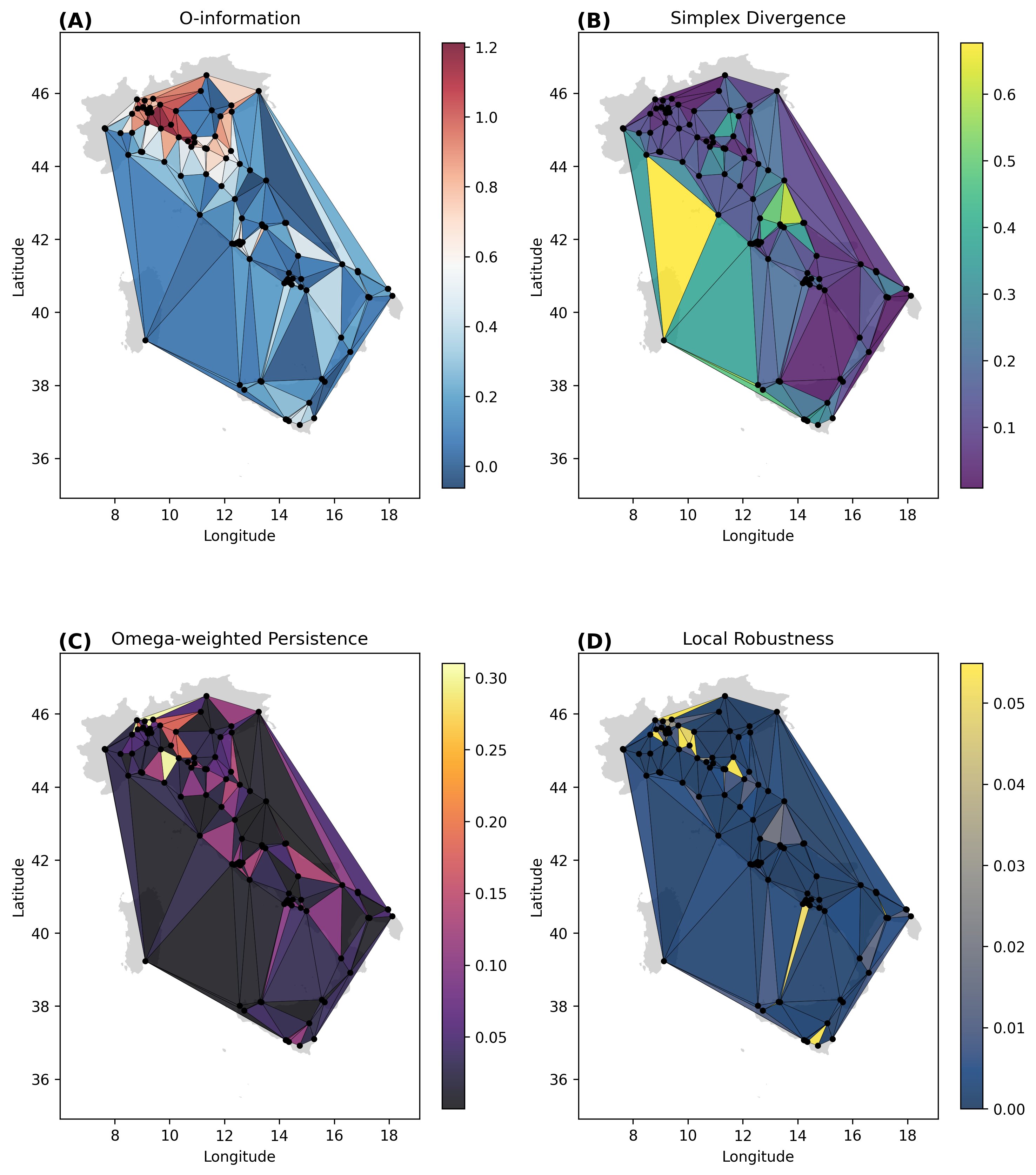}
\end{center}
\caption{Spatial distribution of higher-order interactions and topological descriptors in the Italian air-quality monitoring network. (A) O-information computed from temporally aligned \(\mathrm{NO_2}\)-\(\mathrm{O_3}\) observations at neighboring monitoring stations. (B) Jensen--Shannon simplex divergence describing the heterogeneity of local pollutant-regime distributions. (C) \(\Omega\)-weighted persistence, highlighting topological structures associated with strong higher-order interactions. (D) Local robustness, measuring the spatial coherence of the O-information field with respect to neighboring simplices. Monitoring stations are represented by black dots and the triangulation corresponds to the Delaunay simplicial complex used throughout the analysis.} \label{fig:pollutants}
\end{figure}

The spatial organization of higher-order interactions among monitoring stations is reported in Figure~\ref{fig:pollutants}. Panel~(A) shows the O-information field computed from the aligned \(\mathrm{NO}_2\)-\(\mathrm{O}_3\) trajectories associated with neighboring stations. Most simplices exhibit values close to zero, indicating relatively weak higher-order organization over large portions of the network. Nevertheless, a pronounced positive region emerges in Northern Italy, particularly in the Po Valley area, where O-information reaches its largest values. This behavior reveals the presence of strongly redundant pollutant dynamics, suggesting that neighboring stations belonging to this region share a common atmospheric regime driven by similar emission patterns and meteorological conditions. The simplex divergence map reported in Figure~\ref{fig:pollutants}(B) exhibits a markedly different spatial organization. High divergence values are observed in a limited number of transition regions, whereas most simplices remain characterized by low divergence. Since the divergence is computed from the Jensen--Shannon expression among the probability distributions of the \(\mathrm{NO}_2/\mathrm{O}_3\) regimes observed at neighboring stations, these results indicate that most monitoring sites belong to relatively homogeneous atmospheric environments. The localized high-divergence regions identify areas where different pollutant regimes coexist within the same spatial neighborhood. Figure~\ref{fig:pollutants}(C) presents the \(\Omega\)-weighted persistence field. The strongest values are concentrated in the same northern region highlighted by the O-information map, demonstrating that the most persistent topological structures coincide with the most intense higher-order interactions. This result suggests the existence of a robust large-scale atmospheric organization that remains stable across multiple O-information thresholds. Conversely, most central and southern stations are associated with weaker \(\Omega\)-weighted persistence values, indicating either smaller O-information magnitudes or shorter persistence intervals. The local robustness map shown in Figure~\ref{fig:pollutants}(D) highlights a small number of highly stable simplices embedded within a generally low-robustness background. Robustness maxima are located in localized northern and southern clusters, where large O-information values coexist with limited variability among neighboring simplices. These regions identify spatially coherent pollutant structures that are resilient to local fluctuations of the O-information field. Collectively, the four panels reveal that the strongest pollutant interactions are concentrated within restricted geographical domains rather than being uniformly distributed across the national network.

\begin{figure}[!t]
\begin{center}
\includegraphics[width=0.9\linewidth]{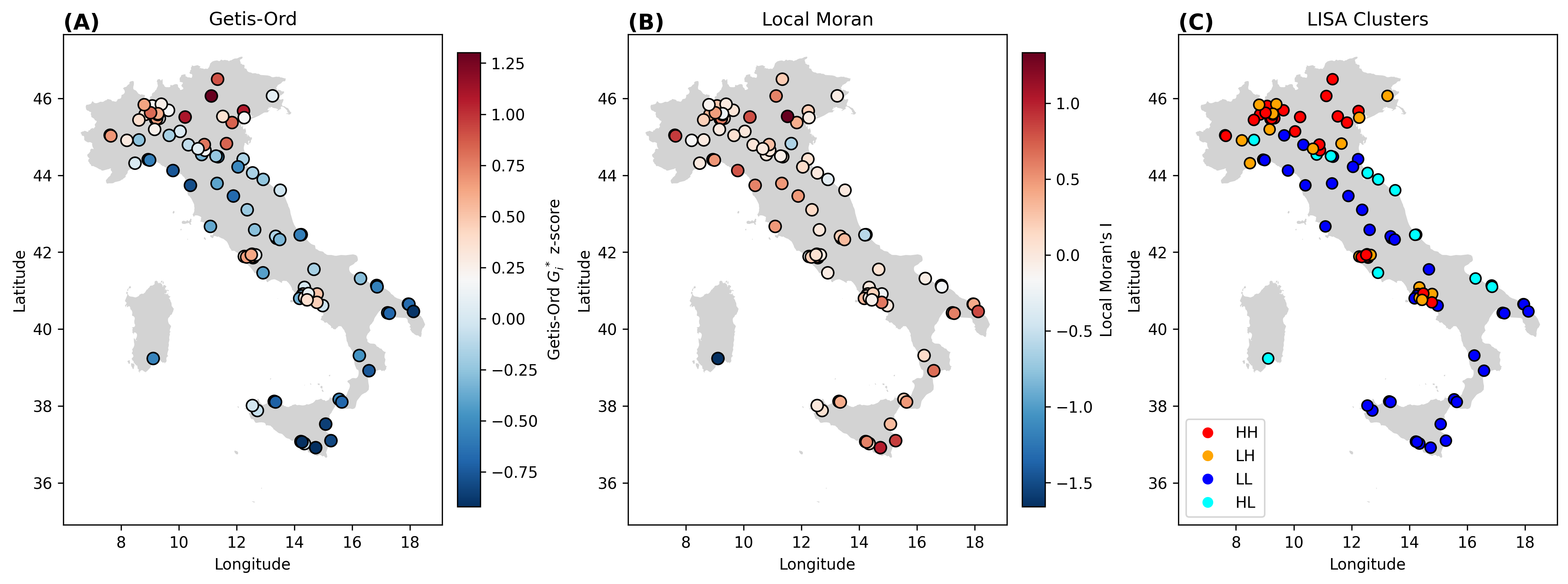}
\end{center}
\caption{Spatial autocorrelation analysis of the station-averaged NO$_2$/O$_3$ ratio across the Italian air-quality monitoring network. (A) Getis--Ord \(G_i^*\) statistic, highlighting local hot-spots and cold-spots of atmospheric regimes through standardized \(z\)-scores. (B) Local Moran's \(I\), quantifying the degree of local spatial autocorrelation and identifying stations embedded in coherent neighborhoods or behaving as spatial outliers. (C) Local Indicators of Spatial Association (LISA) cluster map, where HH (High--High) and LL (Low--Low) denote spatial clusters of similar atmospheric conditions, while HL (High--Low) and LH (Low--High) identify local anomalies relative to their surrounding stations.} \label{fig:moran_getis}
\end{figure}

The Getis--Ord statistics shown in Figure~\ref{fig:moran_getis}(A) identify only weak and spatially diffuse hot- and cold-spot structures, despite the statistically significant positive Moran index obtained for the NO$_2$/O$_3$ ratio. In contrast, the O-information field reported in Figure~\ref{fig:pollutants}(A) exhibits a markedly localized organization, with a pronounced redundant region concentrated over Northern Italy. This result suggests that the higher-order dependencies detected by O-information are not merely a consequence of conventional spatial autocorrelation in the pollutant ratio but emerge from genuinely higher-order interactions among neighboring monitoring stations. A similar distinction can be observed by comparing the LISA clusters in Figure~\ref{fig:moran_getis}(C) with the simplex divergence distribution shown in Figure~\ref{fig:pollutants}(B). While the LISA analysis primarily captures classical spatial clustering patterns of similar atmospheric regimes, the Jensen--Shannon simplex divergence reveals transition regions characterized by heterogeneous pollutant compositions. Consequently, the two quantities provide complementary information: LISA identifies neighborhoods sharing similar environmental conditions, whereas simplex divergence quantifies local heterogeneity among neighboring stations independently of the existence of statistically significant spatial clusters. The comparison becomes even more informative at the topological level. Although the spatial autocorrelation diagnostics reveal moderate clustering of pollutant regimes, neither the Local Moran maps nor the Getis--Ord statistics reproduce the spatial organization highlighted by the \(\Omega\)-weighted persistence field of Figure~\ref{fig:pollutants}(C). The latter specifically emphasizes regions where strong higher-order interactions remain stable across filtration scales, thereby combining information-theoretic intensity and topological robustness. Likewise, the local robustness map reported in Figure~\ref{fig:pollutants}(D) identifies a small number of exceptionally coherent structures that are only partially reflected by the Moran and Getis analyses. Taken together, these observations indicate that the information-topological framework captures organizational properties of the pollutant network that extend beyond standard spatial autocorrelation and hotspot detection methods, revealing multiscale structures associated with higher-order environmental interactions.

\subsection{Multi-hazard Assessment}

\subsubsection{Province of Bari}

\begin{figure}[!t]
\begin{center}
\includegraphics[width=\linewidth]{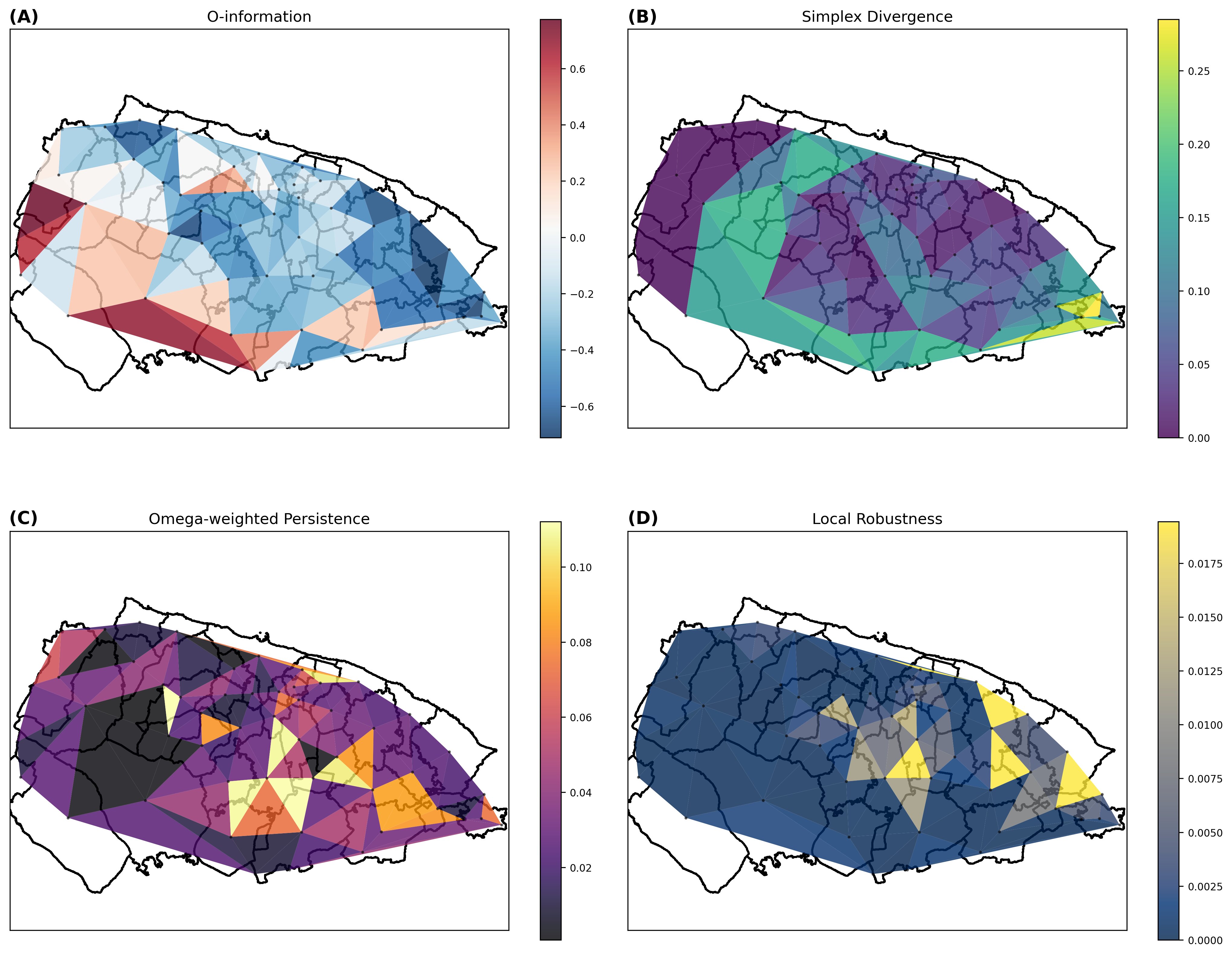}
\end{center}
\caption{Spatial distribution of higher-order interactions and topological descriptors for the multi-hazard assessment of the Province of Bari. (A) O-information highlighting redundant and synergistic multi-hazard interactions among neighboring municipalities. (B) Jensen-Shannon simplex divergence describing the heterogeneity of municipality susceptibility compositions. (C) \(\Omega\)-weighted persistence, identifying persistent topological structures associated with strong higher-order interactions. (D) Local robustness, quantifying the spatial coherence of the O-information field across adjacent simplices. Municipality centroids define the Delaunay simplicial complex, while administrative boundaries are shown in the background.}\label{fig:bari}
\end{figure}

The multi-hazard analysis performed for the Province of Bari is summarized in Figure~\ref{fig:bari}. The O-information field reported in Figure~\ref{fig:bari}(A) displays a heterogeneous mixture of positive and negative values. Positive O-information regions are mainly concentrated in the western and southwestern sectors of the study area, whereas negative values dominate several central and eastern municipalities. This coexistence indicates that both redundancy- and synergy-dominated interactions contribute to the multi-hazard structure of the province. In particular, negative O-information regions suggest that the combined influence of the four susceptibility layers cannot be fully explained through pairwise relationships alone, highlighting the relevance of genuine higher-order interactions. The simplex divergence map in Figure~\ref{fig:bari}(B) is characterized by relatively low values over most of the province, with localized maxima observed along the eastern coastal sector. These results indicate that neighboring municipalities frequently exhibit similar susceptibility compositions, while only a few transition zones display substantial heterogeneity in their susceptibility distributions. The divergence therefore captures spatial variations in multi-hazard composition that are largely independent of the O-information pattern. Figure~\ref{fig:bari}(C) reports the \(\Omega\)-weighted persistence distribution. Several high-intensity structures emerge in the central and eastern sectors of the province. The persistence weighting suppresses regions exhibiting weak O-information or short persistence intervals and emphasizes those simplices characterized by both strong higher-order interactions and stable topological signatures. Consequently, the resulting pattern is considerably more localized than the original O-information field. It is worth noting that the absolute magnitude of the \(\Omega\)-weighted persistence values remains substantially smaller than that observed in the corresponding panel of Figure~\ref{fig:pollutants}, indicating a lower degree of combined interaction strength and topological persistence in the Bari multi-hazard network compared with the national air-quality monitoring network. A complementary view is provided by the local robustness map in Figure~\ref{fig:bari}(D). High robustness values are concentrated in a limited number of isolated clusters, especially along the eastern portion of the province and around selected central municipalities. These regions correspond to areas where O-information varies smoothly among adjacent simplices, indicating the presence of stable multi-hazard configurations. In contrast, the western sector is mostly characterized by low robustness values, suggesting a more fragmented and spatially variable organization of multi-hazard interactions. As for the \(\Omega\)-weighted persistence, the robustness scale is noticeably lower than that observed for the air-quality network in Figure~\ref{fig:pollutants}(D), suggesting that the local coherence of higher-order interactions is weaker and more spatially fragmented in the multi-hazard system.

\subsubsection{Gargano Area}

\begin{figure}[!t]
\begin{center}
\includegraphics[width=\linewidth]{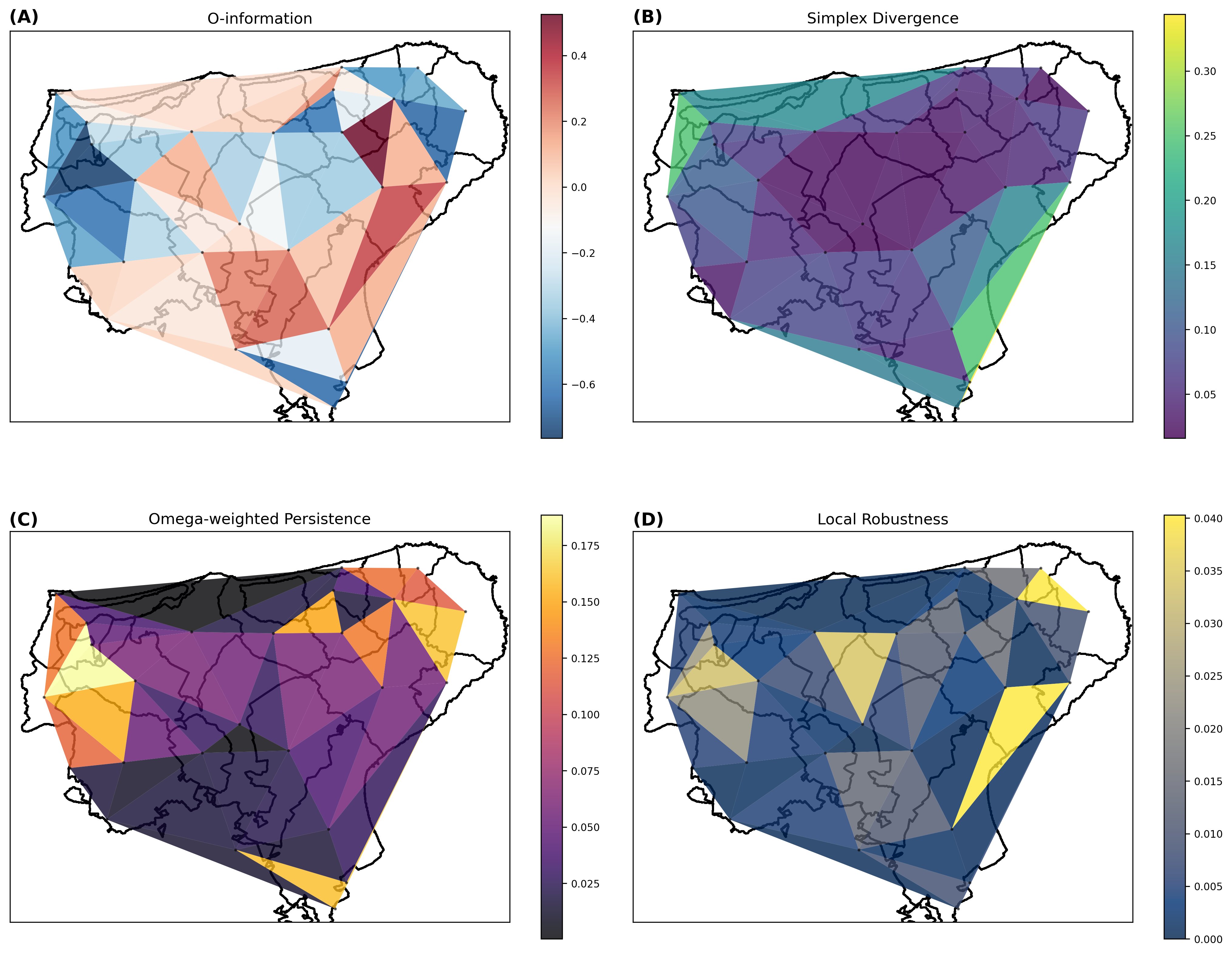}
\end{center}
\caption{Spatial distribution of higher-order interactions and topological descriptors for the multi-hazard assessment of the Gargano area. (A) O-information highlighting redundant and synergistic interactions among neighboring municipalities. (B) Jensen-Shannon simplex divergence quantifying the heterogeneity of local multi-hazard compositions. (C) \(\Omega\)-weighted persistence, emphasizing persistent topological structures associated with strong higher-order interactions. (D) Local robustness, measuring the spatial coherence and stability of the O-information field across adjacent simplices. Municipality centroids are connected through a Delaunay simplicial complex superimposed on the administrative boundaries of the study area.}\label{fig:gargano}
\end{figure}

The results obtained for the Gargano area are shown in Figure~\ref{fig:gargano}. Compared with the Province of Bari, the O-information field exhibits stronger spatial contrasts and a more balanced distribution between synergistic and redundant regimes. Figure~\ref{fig:gargano}(A) reveals extended negative regions in the western sector and positive structures distributed across the central and eastern portions of the peninsula. This pattern indicates the coexistence of distinct multi-hazard interaction mechanisms operating at different spatial scales. The simplex divergence map shown in Figure~\ref{fig:gargano}(B) presents a relatively homogeneous low-divergence core surrounded by more heterogeneous peripheral regions. Divergence maxima are predominantly located along the northwestern and southeastern boundaries, indicating municipalities whose susceptibility compositions differ substantially from those of their neighbors. The central portion of the Gargano is instead characterized by a comparatively coherent multi-hazard structure. The \(\Omega\)-weighted persistence field reported in Figure~\ref{fig:gargano}(C) identifies several highly persistent multi-hazard structures, particularly along the western and eastern sectors of the study area. These regions correspond to simplices exhibiting both strong higher-order interactions and long-lived topological features within the O-information filtration. The concentration of persistent structures near the boundaries of the peninsula suggests the presence of spatially organized multi-hazard regimes associated with distinct environmental conditions. In terms of magnitude, the observed \(\Omega\)-weighted persistence values represent an intermediate condition between those obtained for the national air-quality network (Figure~\ref{fig:pollutants}(C)) and those observed for the Province of Bari (Figure~\ref{fig:bari}(C)), suggesting a level of topological organization that is stronger than that of the Bari multi-hazard landscape while remaining less pronounced than in the pollutant interaction network. Finally, the local robustness map in Figure~\ref{fig:gargano}(D) highlights a small number of highly stable clusters embedded within a broader low-robustness background. Robustness maxima occur in the western and eastern portions of the Gargano, whereas central municipalities generally display intermediate or low values. The correspondence between the robustness field and the \(\Omega\)-weighted persistence distribution indicates that the most persistent topological structures are also associated with locally coherent O-information landscapes. Consistent with the behavior observed for \(\Omega\)-weighted persistence, the robustness values also occupy an intermediate range between the higher levels characterizing the air-quality network and the generally lower levels observed in the Province of Bari. This result suggests that the Gargano area exhibits a moderate degree of spatial coherence in its higher-order interactions, reflecting a balance between the strongly organized structures observed in the pollutant network and the more fragmented multi-hazard organization identified in the Bari case study. The comparison between Figures~\ref{fig:bari} and~\ref{fig:gargano} reveals that the Gargano area shows stronger spatial compartmentalization of multi-hazard interactions than the Province of Bari. While Bari is characterized by relatively smooth variations of O-information and divergence, the Gargano displays more pronounced transitions between synergistic and redundant regimes together with localized regions of elevated persistence and robustness. These findings suggest that topological descriptors are capable of identifying spatially organized multi-hazard structures that remain partially hidden when considering O-information or divergence alone.

\begin{figure}[!t]
\begin{center}
\includegraphics[width=\linewidth]{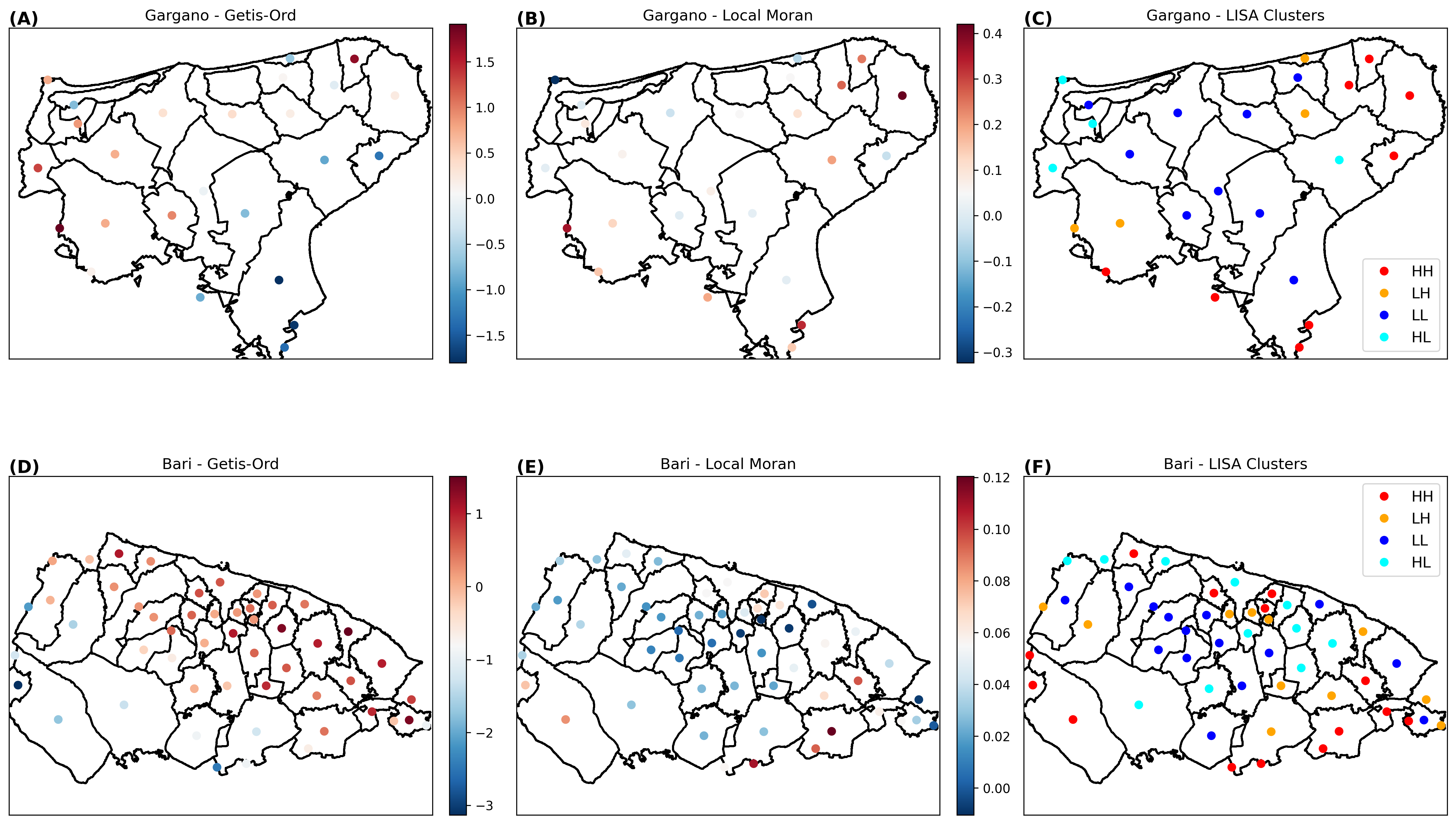}
\end{center}
\caption{Spatial autocorrelation analysis of municipality-level multirisk compositions for the Gargano area (top row) and the Province of Bari (bottom row). Panels (A) and (D) report the multivariate Getis--Ord \(G_i^*\) statistic, identifying municipalities belonging to local hot-spots and cold-spots of multirisk organization. Panels (B) and (E) show the corresponding multivariate Local Moran's \(I\), quantifying the intensity of local spatial autocorrelation among municipality risk compositions. Panels (C) and (F) display the associated Local Indicators of Spatial Association (LISA) clusters, where HH (High--High) and LL (Low--Low) denote municipalities embedded within coherent neighborhoods of similar multirisk composition, while HL (High--Low) and LH (Low--High) identify spatial outliers.}\label{fig:multi_moran_getis}
\end{figure}

The spatial-statistical analyses reported in Figure~\ref{fig:multi_moran_getis} provide an interesting benchmark for interpreting the information-theoretic and topological structures highlighted in Figures~\ref{fig:bari} and \ref{fig:gargano}. In both study areas, Local Moran's \(I\) identifies municipalities embedded within coherent neighborhoods sharing similar multirisk compositions, while the corresponding LISA maps reveal a mixture of HH, LL, HL, and LH configurations. The presence of spatially coherent HH and LL clusters confirms that part of the municipality-level risk organization is already captured by conventional spatial autocorrelation. However, the existence of several HL and LH municipalities, particularly in Gargano, indicates the presence of local transitions and compositional anomalies that cannot be fully described by global clustering patterns alone. The relationship between these spatial clusters and the O-information fields is particularly informative. Regions exhibiting positive O-information often coincide with HH configurations, suggesting that redundancy-dominated multirisk interactions emerge preferentially within spatially homogeneous neighborhoods. Conversely, several negative O-information regions are associated with LISA transition patterns (HL and LH), indicating that synergistic interactions frequently occur near boundaries between municipalities characterized by distinct multirisk compositions. This behavior is especially evident in Gargano, where stronger alternations between synergistic and redundant simplices are observed together with a larger number of local spatial outliers. The comparison with simplex divergence further highlights the complementarity of the two approaches. While Local Moran's \(I\) measures similarity among neighboring municipalities, simplex divergence quantifies differences in the complete probability distributions of risk classes. Consequently, municipalities belonging to the same HH or LL region may still participate in simplices characterized by moderate or elevated divergence values, revealing internal compositional differences that remain invisible to traditional spatial autocorrelation measures. In this sense, Moran's \(I\) captures spatial coherence, whereas simplex divergence characterizes multirisk heterogeneity. The most striking differences emerge when considering \(\Omega\)-weighted persistence and local robustness. Several municipalities identified as significant Getis--Ord hot-spots correspond to regions where \(\Omega\)-weighted persistence assumes relatively large values, indicating that strong local clustering is often accompanied by persistent higher-order interaction structures. Nevertheless, the correspondence is not one-to-one. \(\Omega\)-weighted persistence additionally incorporates the stability of topological features across filtration scales and therefore emphasizes structures that remain relevant over a broad range of interaction strengths. Similarly, local robustness identifies coherent O-information landscapes that are only partially reproduced by Moran or Getis analyses. Municipalities displaying high robustness are not necessarily located within the strongest hot-spots; rather, they correspond to regions where higher-order interactions vary smoothly across neighboring simplices. These comparisons suggest that classical spatial statistics and information-topological descriptors characterize different but complementary properties of multirisk systems. Moran's \(I\), LISA, and Getis--Ord statistics primarily describe the spatial aggregation of similar municipality risk compositions, whereas O-information, simplex divergence, \(\Omega\)-weighted persistence, and local robustness reveal how these compositions interact, organize, and persist across scales. The partial overlap and partial mismatch between the two sets of results indicate that the proposed framework captures organizational patterns extending beyond conventional spatial autocorrelation, providing additional insight into the higher-order structure of environmental risk landscapes.

\begin{table}[t!] 
\centering 
\caption{Summary of persistence-diagram statistics for the three case studies.} \label{tab:tui_summary} \begin{tabular}{lcc}  
Case Study & $m$ & $TUI$ \\ \hline 
Air-Quality Network (Italy) & 125 & 0.88 \\ Multi-hazard Assessment (Bari) & 63 & 0.92 \\ Multi-hazard Assessment (Gargano) & 30 & 0.85 \\ 
\hline 
\end{tabular} 
\end{table}

The comparative indicators reported in Table~\ref{tab:tui_summary} provide a synthetic view of the topological organization of the three investigated systems. The national air-quality network exhibits the largest number of persistence intervals (\(m=125\)), reflecting the greater complexity of the monitoring-station network. In contrast, the Bari and Gargano multi-hazard analyses produce progressively smaller persistence diagrams (\(m=63\) and \(m=30\), respectively). All three case studies are characterized by relatively large values of the \(TUI\), indicating that the persistence spectrum is distributed across multiple competing topological scales rather than being dominated by a small number of exceptionally persistent structures. However, the interpretation of \(TUI\) should be considered with some caution. Since persistence entropy is computed from the normalized persistence lengths \( q_i=\frac{\ell_i}{\sum_j \ell_j}, \) the resulting measure depends exclusively on the relative distribution of persistence intervals and does not explicitly account for the magnitude of the underlying O-information values. Consequently, high persistence entropy identifies a broad distribution of topological scales, but it does not necessarily imply stronger higher-order interactions. In particular, the larger \(TUI\) observed for the Province of Bari (\(TUI=0.92\)) should be interpreted as evidence of a more fragmented persistence landscape rather than as a definitive indication of greater topological complexity or environmental organization. From this perspective, the spatial distributions of \(\Omega\)-weighted persistence reported in Figures~\ref{fig:pollutants}--\ref{fig:gargano} provide a more informative comparison of the three systems, since they jointly incorporate topological persistence and the strength of higher-order interactions. While the Bari case exhibits the largest persistence entropy, its \(\Omega\)-weighted persistence values remain substantially lower than those observed in the air-quality network and generally below those characterizing the most persistent structures of the Gargano area. The air-quality network therefore emerges as the system displaying the strongest combination of interaction intensity and topological uncertainty, whereas the Bari and Gargano landscapes are characterized by weaker but spatially localized persistent structures.

\begin{figure}[!t]
\begin{center}
\includegraphics[width=\linewidth]{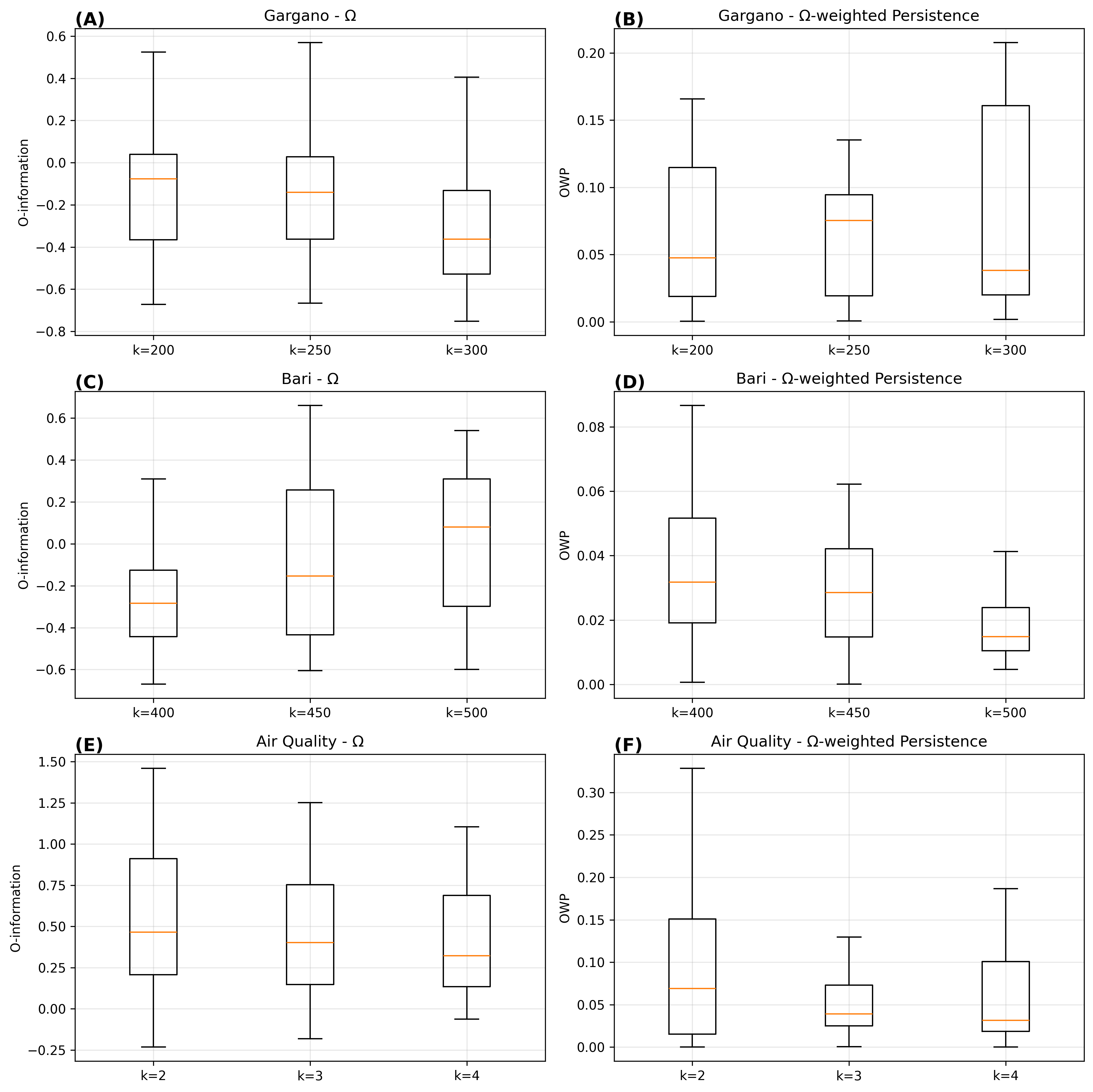}
\end{center}
\caption{Sensitivity analysis of the O-information estimator with respect to the nearest-neighbor parameter \(k\). Panels (A) and (B) show the distributions of O-information and \(\Omega\)-weighted persistence for the Gargano multirisk case (\(k=200,250,300\)), respectively. Panels (C) and (D) report the corresponding distributions for the Province of Bari (\(k=400,450,500\)), while panels (E) and (F) refer to the air-quality monitoring network (\(k=2,3,4\)).}\label{fig:knn_sensitivity}
\end{figure}

Figure~\ref{fig:knn_sensitivity} summarizes the sensitivity of the proposed framework to variations of the nearest-neighbor parameter used in the KSG estimator. The distributions of both O-information and \(\Omega\)-weighted persistence remain relatively stable across the investigated range of \(k\) values, indicating that the main conclusions of the analysis are not critically dependent on a specific parameter choice. For the Gargano case study (Figures~\ref{fig:knn_sensitivity}A--B), the O-information distributions preserve their predominantly negative character for all tested values of \(k\), with comparable interquartile ranges and only moderate shifts in the median. This behaviour suggests that the synergy-dominated organization observed in the multirisk landscape is robust with respect to the estimator configuration. The corresponding \(\Omega\)-weighted persistence distributions exhibit slightly larger variability, particularly for \(k=300\), but maintain a similar overall range and central tendency, indicating that the identification of persistent higher-order structures is not strongly affected by moderate variations of the neighborhood size. The Province of Bari (Figures~\ref{fig:knn_sensitivity}C--D) displays a similarly stable behaviour. Although the median O-information progressively shifts toward less negative values as \(k\) increases, the overall shape of the distributions remains remarkably consistent. The associated \(\Omega\)-weighted persistence values decrease slightly for larger \(k\), reflecting the smoothing effect commonly observed in nearest-neighbor estimators. Nevertheless, the distributions preserve comparable dispersion and do not exhibit abrupt structural changes, suggesting that the detected topological patterns represent intrinsic properties of the multirisk system rather than artifacts of a particular parameter choice. The air-quality network (Figures~\ref{fig:knn_sensitivity}E--F) shows the greatest stability among the three applications. The distributions of O-information remain consistently positive, indicating a persistent redundancy-dominated organization of pollutant dynamics across all tested values of \(k\). While the median O-information decreases gradually from \(k=2\) to \(k=4\), the overall distributional structure is preserved. An analogous behaviour is observed for \(\Omega\)-weighted persistence, whose central tendency and spread remain within a relatively narrow interval despite the parameter variation. These results indicate that the higher-order interaction patterns identified in the three case studies are robust with respect to the selection of the KSG nearest-neighbor parameter. Variations of \(k\) primarily affect the numerical magnitude of the estimated quantities, whereas the qualitative organization of the distributions, the dominance of synergistic or redundant regimes, and the persistence of the associated topological structures remain substantially unchanged. This evidence supports the stability of the proposed information-topological framework and suggests that the reported environmental patterns are not artifacts of a specific estimator configuration.

\section{Discussion}

The present work introduces a unified framework combining O-information, persistent homology, simplex divergence, and local robustness to characterize the spatial organization of environmental systems. Unlike conventional geostatistical or network-based approaches, which primarily focus on pairwise relationships among neighboring locations, the proposed methodology explicitly accounts for higher-order interactions among triplets of spatial entities and investigates their topological uncertainty across multiple scales. 

An important methodological aspect of the proposed framework is that the topology of the spatial system is not inferred from the environmental variables themselves, but from a fixed Delaunay representation of geographical proximity. Consequently, the persistent structures identified in the filtration reflect the organization of the O-information field across scales rather than changes in the underlying neighborhood geometry. This separation between spatial support and filtration function improves interpretability and reduces the sensitivity of the analysis to small geometric perturbations.

The analysis of the air-quality monitoring network reveals the existence of localized regions characterized by strong positive O-information, particularly in Northern Italy. Such structures indicate redundancy-dominated interactions among neighboring stations, suggesting the presence of common atmospheric drivers acting across multiple locations. This observation is consistent with previous studies showing that pollutant dynamics are frequently controlled by large-scale meteorological forcing, emission corridors, and regional transport processes (\cite{monks2015tropospheric, seinfeld2016atmospheric}). However, most existing air-quality analyses rely on correlations, covariance structures, or graph-theoretical measures computed on pairwise relationships. By contrast, O-information directly captures higher-order interactions among neighboring stations and therefore provides information that cannot be recovered from standard correlation-based frameworks. 

The simplex divergence results further suggest that the strongest O-information structures do not necessarily coincide with the most heterogeneous atmospheric regimes. This distinction highlights an important conceptual point. Regions characterized by strong higher-order interactions may remain relatively homogeneous in terms of pollutant-regime composition, whereas high simplex divergence identifies transition zones where different atmospheric conditions coexist. In this sense, O-information and divergence capture complementary aspects of the same environmental system: interaction strength versus configurational diversity. 

From a topological perspective, the persistence analysis reveals that only a limited subset of simplices contributes substantially to the persistent organization of the pollutant network. Similar observations have emerged in recent applications of topological data analysis to climate and environmental systems, where persistent topological structures have been associated with robust spatial patterns and coherent environmental regimes (\cite{topology_climate}). The present results extend these ideas by linking persistence directly to higher-order information interactions rather than to geometric or correlation-based quantities alone. The resulting \(\Omega\)-weighted persistence therefore identifies spatial structures that are simultaneously information-rich and topologically persistent. 

The multi-hazard analyses of the Province of Bari and the Gargano area exhibit a markedly different behavior. In both cases, positive and negative O-information values coexist within the same spatial domain, indicating the simultaneous presence of redundant and synergistic multi-hazard configurations. Synergistic regions are particularly interesting because they suggest that the combined behavior of multiple hazards cannot be inferred from pairwise interactions alone. This finding is consistent with the growing multi-hazard literature emphasizing the importance of compound and cascading hazards (\cite{gill2016review,zscheischler2020future}). Traditional multi-hazard indicators are typically constructed through weighted linear combinations of hazard maps or multicriteria decision analyses. While these approaches provide useful aggregate measures, they often neglect the possibility of higher-order dependencies among susceptibilities. The present framework addresses this limitation by evaluating directly the information shared among multiple neighboring municipalities. The application of nearest-neighbor entropy estimators to categorical environmental variables requires particular care because repeated observations may generate ties and zero-distance neighborhoods. In the present study, however, municipalities are represented by ensembles of multivariate risk vectors sampled at the pixel level rather than by single aggregated indicators. Consequently, the estimation procedure operates on empirical multivariate distributions rather than on isolated categorical observations. We acknowledge that KSG-based estimators were originally formulated for continuous variables, nevertheless, purely discrete estimators would face a different limitation in the present context: considering four risk variables and four classes per variable, the configuration space contains \(4^4=256\) possible states, leading to severe sparsity and numerous poorly sampled or empty configurations at the municipality level. For this reason, the nearest-neighbor formulation was adopted as a practical compromise for characterizing multivariate dependencies in sparse environmental distributions. A systematic comparison between nearest-neighbor estimators and bias-corrected discrete entropy estimators remains an important direction for future work.

The adoption of Jensen-Shannon simplex divergence further improves the characterization of multi-hazard landscapes. Previous approaches in susceptibility assessment generally rely on municipality-level averages or dominant classes (\cite{Malczewski01082006,carpignano2009multi,kappes2012}). In contrast, the present methodology exploits the complete distribution of susceptibility labels within each municipality and quantifies differences among neighboring municipalities through a Jensen--Shannon formulation. Consequently, spatially coherent regions with similar dominant susceptibilities but different internal compositions remain distinguishable, providing a more sensitive description of multi-hazard transitions. 

The persistence results reveal substantial differences between the Bari and Gargano case studies. Although both areas display comparable ranges of O-information, the Gargano dataset is characterized by more spatially localized persistent structures and stronger compartmentalization of higher-order interactions. This behavior likely reflects the more heterogeneous geomorphological and environmental setting of the Gargano peninsula, where sharp transitions between coastal, agricultural, and inland mountainous systems generate distinct multi-hazard regimes. Conversely, the Province of Bari exhibits a more diffuse topological organization associated with a larger diversity of competing persistence scales. The proposed interpretation remains to be validated through additional analyses and will be explored in future developments of this study.

The persistence entropy values obtained for all case studies indicate that the corresponding persistence spectra are distributed across multiple topological scales, with normalized Topological Uncertainty Index values consistently larger than 0.85. However, since persistence entropy depends only on the relative distribution of persistence lengths, it should be interpreted primarily as a measure of the dispersion or fragmentation of the persistence landscape rather than as a direct quantification of uncertainty. In this respect, the larger value observed for the Province of Bari (\(TUI=0.92\)) suggests a broader distribution of competing topological scales, but does not necessarily imply stronger or more structured higher-order interactions. A more complete picture emerges when persistence length is combined with O-information through the \(\Omega\)-weighted persistence measure. From this perspective, the air-quality network exhibits substantially larger values than both multi-hazard case studies, indicating that its topological structures are supported by stronger higher-order interactions and remain stable across a wider range of filtration thresholds. These results suggest that persistence entropy and \(\Omega\)-weighted persistence characterize complementary aspects of the system, namely the diversity of topological scales and the strength of the corresponding interaction patterns.

The main computational costs arise from nearest-neighbor O-information estimation and persistent homology. In the present applications, the use of two-dimensional Delaunay triangulations keeps the complex sparse and restricts O-information to local triplets. Larger national or continental datasets may require multi-resolution sampling, sparse or landmark-based filtrations, and parallel implementations.

A further contribution of this work is the introduction of a local robustness measure based on spatial variations of the O-information field. Unlike persistence entropy, which provides a global characterization of topological complexity, local robustness identifies coherent neighborhoods where strong higher-order interactions remain stable with respect to neighboring simplices. The complementarity between local robustness and persistence entropy allows the simultaneous investigation of local and global scales, an aspect rarely considered in current applications of topological data analysis to environmental systems. 

To the best of our knowledge, previous studies have independently employed O-information to investigate redundancy and synergy (\cite{rosas2019oinfo}) and persistent homology to characterize spatial structures (\cite{Rote2006, carlsson2009topology, edelsbrunner2010computational, otter2017roadmap}). However, a systematic integration of O-information, simplicial complexes, persistence-based descriptors, and topological uncertainty for environmental and multi-hazard systems has not been previously reported. The proposed framework therefore establishes a direct connection between information theory and topological data analysis, providing a new interpretation of environmental uncertainty as the instability of higher-order spatial organization across interaction scales. 

Although persistence diagrams provide a robust multiscale representation of the O-information field, a complete statistical assessment of the resulting structures requires the specification of suitable null models. In the present framework, the construction of such models is nontrivial because municipalities are represented by multivariate empirical risk distributions and the associated O-information depends on both marginal probabilities and higher-order dependency structures. Consequently, different randomization schemes, including label shuffling, bootstrap resampling, subsampling, or correlation-preserving surrogates, correspond to distinct null hypotheses. The development and comparison of these null-model families represents an important direction for future work. Accordingly, the present study should be interpreted as an exploratory methodological investigation aimed at introducing and evaluating the proposed information-topological framework, while a systematic analysis of statistical significance and uncertainty remains a topic for subsequent research.

Future developments may include the use of persistent cohomology representatives to establish a more rigorous localization of topological features, the extension to temporal filtrations for evolving environmental systems, and the incorporation of additional information-theoretic measures such as integrated information decomposition or multivariate transfer entropy. The persistent homology analysis is constructed from the filtration function \(f(\tau)=|\Omega(\tau)|\), where the absolute value of the O-information quantifies the strength of higher-order organization independently of its nature. This choice allows redundancy-dominated (\(\Omega>0\)) and synergy-dominated (\(\Omega<0\)) interactions to be treated on a common non-negative scale measuring the departure from statistical independence. While the sign information is therefore not retained within the filtration itself, it remains fully available in the O-information maps and in the interpretation of the detected structures. An interesting extension of the framework would consist in constructing separate filtrations for positive and negative O-information values, thereby allowing the topological organization of redundancy- and synergy-dominated structures to be investigated independently. These directions may provide a more comprehensive understanding of how higher-order interactions shape the topology of complex environmental networks.





\section*{Conflict of Interest Statement}


The authors declare no competing interests.

\section*{Author Contributions}


Conceptualization: DP, NA; Software: DP; Data Curation: AF, GRS, AC, RN; Methodology: DP; Formal analysis and investigation: DP; Visualization: DP; Funding acquisition: RB, SS, AM, NA, EP; Resources: DP, GRS, RN, EP, AM, NA; Supervision: DP, AM, NA; Writing - original draft preparation: DP, AF, GRS, AC; Writing - review and editing: DP, AF, GRS, LB, DC, RC, AC, DD, ME, FG, NK, RL, RN, EP, ST, RB, SS, AM, NA.

\section*{Funding}

Authors were supported by the Italian funding within the “Budget MUR - Dipartimenti di Eccellenza 2023 - 2027” (Law 232, 11 December 2016) - Quantum Sensing and Modelling for One-Health (QuaSiModO), CUP:H97G23000100001. This paper was funded by the Italian Ministry of University and Research (MUR) within the framework of the National Programme “Research, Innovation and Competitiveness for the Green and Digital Transition” (PN RIC) 2021–2027, Call D.D. No. 307 – Project ECHO TWIN. The project is co-funded by the European Union through the European Regional Development Fund (ERDF). Azione 1.1.2$\_$CUP: B99H26000290007; Azione 1.1.3b$\_$CUP: B92F26000440005; Azione 1.4.3$\_$CUP: B89J26003490005. Funding to Università degli Studi di Bari: Azione 1.1.2$\_$CUP: B99H26000520007; Azione 1.1.3b$\_$CUP: B92F2600053000; Azione 1.4.3$\_$CUP: B99J26001800005.



\section*{Data Availability Statement}

The air-quality dataset analyzed for this study can be found using the package in the Github repository \url{https://github.com/PaoloMaranzano/EEAaq_R}. Any additional dataset and code are available from the corresponding author upon reasonable request.

\bibliography{apssamp}

\end{document}